\documentclass[11pt]{article}

\usepackage[preprint]{acl}

\usepackage{times}
\usepackage{latexsym}

\usepackage[T1]{fontenc}

\usepackage[utf8]{inputenc}

\usepackage{CJKutf8}

\usepackage{microtype}

\usepackage{graphicx}

\usepackage{booktabs}
\usepackage{multirow}
\usepackage{array}

\usepackage{algorithm}
\usepackage{algorithmic}

\usepackage{amsmath}
\usepackage{amssymb}

\usepackage{xcolor}
\usepackage{soul} 
\usepackage{pifont} 
\usepackage{colortbl}

\usepackage{tikz}
\usetikzlibrary{positioning}
\usepackage{pgfplots}
\pgfplotsset{compat=1.17}

\usepackage{placeins}

\usepackage{enumitem}

\usepackage{tcolorbox}

\title{Complexity-Aware Training for OOD Generalization in Dense Retrieval}

\author{
  Xincan Feng\textsuperscript{1,2,\dag} \quad
  Noriki Nishida\textsuperscript{2} \quad
  Yusuke Sakai\textsuperscript{1} \quad
  Yuji Matsumoto\textsuperscript{2} \\[0.5em]
  \textsuperscript{1}Nara Institute of Science and Technology \\
  \textsuperscript{2}RIKEN Center for Advanced Intelligence Project \\
  \textsuperscript{\dag}\texttt{cecilia.fung@foxmail.com}
}

\begin{document}
\maketitle

\begin{abstract}
Synthetic query generation has become essential for training dense retrievers, yet prior methods generate one query per document, focusing solely on query quality. We are the first to systematically study multi-query synthesis and discover a \textbf{quality-diversity trade-off}: high-quality queries benefit in-domain tasks, while diverse queries benefit out-of-domain (OOD) generalization. Through controlled experiments on 4 benchmark types across Contriever, RetroMAE, and Qwen3-Embedding, we find that diversity benefit strongly correlates with query complexity ($r$$\geq$0.95, $p$$<$0.05), approximated by content words (CW). We formalize this as the \textbf{Complexity-Diversity Principle (CDP)}: \textit{query complexity determines optimal diversity}. Based on CDP, we propose \textbf{complexity-aware training}: multi-query synthesis for high-complexity tasks and CW-weighted training for existing data. Both strategies improve OOD performance on reasoning-intensive benchmarks, with compounded gains when combined.
\end{abstract}

\section{Introduction}

Dense retrieval models encode queries and documents into a shared embedding space for efficient similarity search \citep{karpukhin2020dense,xiong2021approximate}. Training these models requires large-scale query-document pairs, which are expensive to annotate manually. \textit{Synthetic query generation} addresses this by using LLMs to generate training queries from documents \citep{bonifacio2022inpars,dai2022promptagator}, aiming for retrievers that generalize to unseen tasks.

Existing methods generate \textbf{one query per document}, focusing on query quality. InPars \citep{bonifacio2022inpars}, Promptagator \citep{dai2022promptagator}, and SAP \citep{thakur2024swimir} use few-shot prompting to produce high-quality queries resembling human-written ones. DRAGON \citep{lin2023dragon} combines multiple data sources but still produces one query per document, where its ``diversity'' refers to data source variety rather than per-document query diversity. \textbf{No prior work studies generating multiple queries for the same document.} This leaves a fundamental question unexplored: \textit{does multi-query diversity help retrieval, and if so, when?}

\begin{figure}[!t]
\centering
\small
\begin{tcolorbox}[
  width=0.95\columnwidth,
  colback=white,
  colframe=gray!40,
  colbacktitle=gray!20,
  coltitle=black,
  title=\textbf{Query:} What does \ul{Ivan} \ul{promise} to do when he \ul{turns} \ul{thirty}?,
  fonttitle=\small,
  boxrule=0.5pt,
  arc=4pt,
  left=4pt, right=4pt, top=4pt, bottom=4pt
]
\textbf{Q/Doc=1} \textcolor{red}{\ding{55}} \textit{``I do love you, Ivan. Dmitri says of you---Ivan is a tomb!''} \\
\textcolor{gray}{$\hookrightarrow$ Matches character name only} \\[0.5em]
\textbf{Q/Doc=3} \textcolor{green!50!black}{\ding{51}} \textit{``Listen, Alyosha,'' Ivan began, ``if I am really able to care for the sticky little leaves...''} \\
\textcolor{gray}{$\hookrightarrow$ Ivan's actual promise about living until thirty}
\end{tcolorbox}
\caption{Diverse queries act as implicit regularization. \ul{Underlined} words are content words (CW). Q/Doc=1 overfits to surface features (character names); Q/Doc=3 learns semantic matching and retrieves Ivan's \textit{actual promise} (Contriever on NovelHopQA).}
\label{fig:teaser}
\end{figure}

\begin{figure*}[t]
\centering
\resizebox{\textwidth}{!}{%
\begin{tikzpicture}[
    node distance=0.5cm,
    mainbox/.style={rectangle, draw, rounded corners, minimum width=2.8cm, minimum height=1.8cm, align=center, font=\small, line width=0.8pt},
    smallbox/.style={rectangle, draw, rounded corners=2pt, minimum width=2.2cm, minimum height=0.6cm, align=center, font=\scriptsize},
    inputbox/.style={rectangle, draw, dashed, rounded corners, minimum width=2.0cm, minimum height=0.9cm, align=center, font=\small, gray},
    arrow/.style={->, >=stealth, very thick, line width=1.2pt},
    note/.style={font=\scriptsize, align=center}
]

\node[inputbox] (input) {Documents\\$\mathcal{D} = \{d_1, ..., d_N\}$};

\node[mainbox, fill=blue!15, right=0.8cm of input] (step1) {\textbf{Step 1:}\\Zero-shot Multi-Query\\Synthesis\\$\{q^{(m)}\}_{m=1}^M = \text{LLM}(d, p)$};

\node[mainbox, fill=purple!15, right=1.5cm of step1] (step2) {\textbf{Step 2:}\\Q-D Metrics\\Measurement\\[0.2em]\scriptsize\textbf{Quality}: Dist-Sim, Len-Sim\\\scriptsize\textbf{Diversity}: CE, Self-BLEU};

\node[mainbox, fill=gray!15, right=1.5cm of step2] (step4) {Contrastive\\Training\\$\mathcal{L} = -\log\frac{e^{s(q,d^+)}}{{\sum_j} e^{s(q,d_j)}}$};

\node[inputbox, right=0.8cm of step4] (output) {Retriever\\$f_\theta$};

\node[mainbox, fill=yellow!20, below=0.8cm of step1, minimum width=3.6cm, xshift=2.4cm] (step3) {\textbf{Step 3:}\\Diversity Tuning\\[0.2em]\scriptsize Dist-Sim$\uparrow$, Len-Sim$\uparrow$\\\scriptsize CE$\downarrow$, Self-BLEU$\downarrow$};

\node[mainbox, fill=green!15, below=0.8cm of step4, minimum width=3.2cm] (cwweight) {\textbf{Step 4:}\\CW-Weighting\\[0.2em]\scriptsize $w(q) = \frac{\min(\text{CW}(q), \kappa)}{\sum \min(\text{CW}, \kappa)} \cdot |B|$};

\draw[arrow] (input) -- (step1);
\draw[arrow] (step1) -- (step2);
\draw[arrow] (step2) -- (step4);
\draw[arrow] (step4) -- (output);

\draw[arrow, dashed, orange!70!black, line width=1.5pt] (step2.south) |- (step3.east);
\draw[arrow, dashed, orange!70!black, line width=1.5pt] (step3.west) -| (step1.south);

\node[below=0.2cm of step3, font=\small\bfseries, align=center, text width=4.0cm, text=orange!70!black] {Multi-query as regularization};

\draw[arrow, green!60!black, line width=1.5pt] (cwweight.north) -- (step4.south);

\node[below=0.2cm of cwweight, font=\small\bfseries, align=center, text width=5.0cm, text=green!60!black] {Content words as\\query weighting};

\end{tikzpicture}%
}
\caption{Pipeline. Given a document corpus, we (1) generate $M$ diverse queries per document using zero-shot prompting, (2) measure query quality and diversity using Q-D metrics, (3) tune diversity level based on target task, and (4) apply CW-weighting for complexity-based sample weights during training. Both multi-query and CW-weighting improve OOD generalization.}
\label{fig:method_overview}
\end{figure*}

\begin{table*}[t]
\centering
\scriptsize
\begin{tcolorbox}[
  width=0.98\textwidth,
  colback=white,
  colframe=gray!40,
  colbacktitle=gray!15,
  coltitle=black,
  title=\textbf{Document:} \textit{Results-Based Accountability (RBA) is a disciplined way of thinking and taking action that communities can use to improve the lives of children, youth, families, adults and the community as a whole. RBA is also used by organizations to improve the performance of their programs...},
  fonttitle=\scriptsize,
  boxrule=0.5pt,
  arc=4pt,
  left=4pt, right=4pt, top=4pt, bottom=4pt
]

\renewcommand{\arraystretch}{1.1}
\begin{tabular}{p{0.30\textwidth}p{0.64\textwidth}}
\cellcolor{green!10}\textbf{Supervised Query} & \cellcolor{red!10}\textbf{Few-shot (InPars-GBQ, 3 queries)} \\
\midrule
what is rba &
1. What is RBA and how does it help communities? \newline
2. What is RBA and how is it used to improve community well-being? \newline
3. What is RBA and how does it benefit communities? \newline
\textcolor{gray}{$\hookrightarrow$ All ask \textit{the same question} with minor variations} \\
\end{tabular}

\vspace{0.5em}
\begin{tabular}{p{0.44\textwidth}p{0.50\textwidth}}
\multicolumn{2}{c}{\cellcolor{blue!10}\textbf{Zero-shot Diverse (Ours, 20 queries)}} \\
\midrule
1. What is Results-Based Accountability (RBA)? & 11. Community impact \\
2. What are the main goals of RBA? & 12. Conditions of well-being \\
3. What types of communities can benefit from RBA? & 13. Performance improvement \\
4. How does RBA improve community well-being? & 14. RBA is used to enhance community well-being. \\
5. How can organizations implement RBA in their programs? & 15. RBA helps organizations improve program performance. \\
6. How do leaders measure community impact using RBA? & 16. Leaders work collectively to achieve community impact. \\
7. Why is RBA important for improving lives? & 17. Which groups can utilize RBA for improvement? \\
8. Why do communities need to focus on conditions of well-being? & 18. Is it true that RBA focuses on children and families? \\
9. When should a community start using RBA? & 19. How does RBA compare to traditional methods? \\
10. If a community adopts RBA, what changes can be expected? & 20. What are the differences between community impact and program performance? \\
\multicolumn{2}{l}{\textcolor{gray}{$\hookrightarrow$ Diverse formats: What/How/Why/When/If questions, keywords, statements, comparisons}} \\
\end{tabular}
\end{tcolorbox}
\caption{Comparison of queries generated by different methods for the first document in MS MARCO. Supervised queries are short and keyword-focused. Few-shot methods produce paraphrases of a single pattern. Our zero-shot diverse method generates semantically varied queries covering different aspects and formats.}
\label{tab:fewshot_vs_zeroshot}
\end{table*}

We hypothesize that \textbf{query complexity} determines when diversity helps. We approximate complexity via \textbf{content words (CW)}, i.e., the count of unique non-stopwords in a query (Figure~\ref{fig:teaser}). If true, diversity benefit should correlate with CW. Our contributions are:
\begin{itemize}[leftmargin=1.5em, itemsep=0.2em]
    \item \textbf{Quality-diversity trade-off}: We systematically study multi-query synthesis for the first time, revealing that quality benefits in-domain while diversity benefits OOD.
    \item \textbf{Complexity-Diversity Principle}: Through controlled experiments on 4 benchmark types with Contriever, RetroMAE, and Qwen3-Embedding, we discover that query complexity determines optimal diversity ($r$$\geq$0.95, $p$$<$0.05 in 12/14 conditions), suggesting that high-complexity tasks, e.g., CW$>$10, benefit from diversity-aware training.
    \item \textbf{Complexity-aware training}: Based on CDP, we show that both multi-query synthesis and CW-weighted training improve OOD for high-complexity tasks, with compounded gains when combined.
\end{itemize}

\section{Related Work}

\paragraph{Query Synthesis for Dense Retrieval}
Query synthesis has followed the one-query-per-document paradigm. InPars \citep{bonifacio2022inpars} introduced LLM-based few-shot prompting. Promptagator \citep{dai2022promptagator} improved quality via consistency filtering. SAP \citep{thakur2024swimir} designed summarize-then-ask prompts for multilingual retrieval. DRAGON \citep{lin2023dragon} reduced costs by mixing sentence cropping with docT5query. DRAMA \citep{drama2025} improved few-shot example selection. ReasonEmbed \citep{reasonembed2025} augmented queries with reasoning chains for reasoning-intensive tasks. We explore multiple queries per document and discover the \textbf{quality-diversity trade-off}.

\paragraph{Data Augmentation and OOD Generalization}
Contriever \citep{izacard2022unsupervised} and RetroMAE \citep{xiao2022retromae} use unsupervised pretraining. DPR \citep{karpukhin2020dense} and ANCE \citep{xiong2021approximate} improve hard negative sampling. BEIR \citep{thakur2021beir} benchmarks OOD generalization. While prior work explores pretraining and data scaling, the role of \textit{query diversity} remains unstudied. Our work fills this gap.

\section{Methodology}

To test our hypothesis, we design a controlled experimental framework using \textbf{zero-shot multi-query synthesis}, which enables systematic manipulation of quality and diversity via prompts (Figure~\ref{fig:method_overview}; pseudocode in Appendix~\ref{sec:appendix_algorithm}).

\subsection{Step 1: Zero-shot Multi-Query Synthesis}

Unlike few-shot methods that produce homogeneous queries due to pattern copying (CE$>$0.5; analysis in Appendix~\ref{sec:appendix_fewshot_limitations}), we use zero-shot prompting to generate $M$ semantically diverse queries per document in a single LLM call with temperature=0. The prompt specifies multiple formats: factual, procedural, causal, conditional, keyword, and comparison queries (Table~\ref{tab:fewshot_vs_zeroshot}; full templates in Appendix~\ref{sec:appendix_prompts}). This achieves true semantic diversity, reproducibility, and cost efficiency.

\subsection{Step 2: Q-D Metrics Measurement}

We propose Quality-Diversity (Q-D) metrics to quantify the characteristics of generated queries. For \textit{quality}, we measure how closely synthetic queries resemble human-written ones using \textbf{Dist-Sim}, i.e., embedding similarity, and \textbf{Len-Sim}, i.e., length similarity. For \textit{diversity}, we measure semantic variation among queries using \textbf{CE}, i.e., cross-encoder paraphrase ratio, and \textbf{Self-BLEU}, i.e., lexical overlap. Formal definitions in Appendix~\ref{sec:appendix_qd}.

\subsection{Step 3: Diversity Tuning}

To isolate the effect of diversity, we design two prompt variants with the same query quantity $M$: (1) \textbf{Diverse mode} generates semantically distinct queries using different formats, achieving CE$\approx$0.04; (2) \textbf{Paraphrase mode} generates semantically equivalent queries with surface-level variations, achieving CE$\approx$0.50, serving as a low-diversity control. Full templates in Appendix~\ref{sec:appendix_prompts}.

\subsection{Step 4: CW-Weighted Training}
\label{sec:methodology_cw}

We propose \textbf{CW-weighted training} as a complementary method to multi-query synthesis. While not a perfect measure, content words (CW) provide a simple and practical approximation of query complexity by counting unique non-stopwords:
\begin{equation}
\text{CW}(q) = |\{w \in \text{tok}(q) : w \notin \mathcal{S} \land |w| > 1\}|
\label{eq:cw_method}
\end{equation}
Unlike reasoning-based metrics that require model inference, CW can be computed from queries alone via simple word counting. During training, we apply truncation and batch normalization:
\begin{equation}
w(q) = \frac{\min(\text{CW}(q), \kappa)}{\sum_{q_i \in B} \min(\text{CW}(q_i), \kappa)} \cdot |B|,
\label{eq:cw_weight_method}
\end{equation}
where $\kappa$=100 truncates extreme values and $|B|$ is batch size. CW-weighting can be applied to both single-query and multi-query data for compounded OOD gains.

\section{Experimental Setup}
\label{sec:experiments}

\paragraph{Training Data.}
We use 80k MS MARCO QA v1.1 \citep{nguyen2016msmarco} documents. For each document, we generate synthetic queries using different methods and train dense retrievers.

\paragraph{Query Generator.}
We use GPT-4o-mini \citep{openai2024gpt4o} for main experiments, and Mistral-7B-Instruct to validate on open-source LLMs.

\paragraph{Training Retriever.}
We fine-tune Contriever \citep{izacard2022unsupervised} and RetroMAE \citep{xiao2022retromae} as BERT-based retrievers (batch size 128, 100 epochs), and Qwen3-Embedding-0.6B as an LLM-based retriever (batch size 32, 20 epochs). Model selection is based on val NDCG@10.

\paragraph{Baselines.}
We compare against representative query synthesis methods. \textbf{Supervised} uses human-annotated MS MARCO queries as a reference. \textbf{InPars-GBQ} \citep{bonifacio2022inpars} employs 3-shot prompting with the Guided-by-Bad-Questions variant. \textbf{DRAGON-S} \citep{lin2023dragon} uses sentence cropping with cross-encoder reranking. \textbf{SAP} \citep{thakur2024swimir} applies 5-shot summarize-then-ask prompting. \textbf{DRAMA} \citep{drama2025} is a recent baseline using similar few-shot methods. Training details in Appendix~\ref{sec:appendix_implementation}.

\paragraph{Evaluation Benchmarks.}
We evaluate on four benchmark categories using NDCG@10. Statistics in Appendix~\ref{sec:appendix_datasets}. (1) \textbf{TREC-DL}, i.e., Deep Learning 2019/2020 tracks, serves as in-domain evaluation on MS MARCO passages. (2) \textbf{BEIR} \citep{thakur2021beir} contains 14 publicly available datasets for standard OOD evaluation covering factoid QA, argument retrieval, and scientific domains. (3) \textbf{BRIGHT} \citep{bright2024} contains 12 datasets for reasoning-intensive retrieval requiring complex inference. (4) \textbf{Multi-hop} includes 2WikiMultihopQA \citep{ho2020wikimultihop}, MuSiQue \citep{trivedi2022musique}, HotpotQA \citep{yang2018hotpotqa}, and NovelHopQA \citep{novelhopqa2025}, which require aggregating evidence across multiple documents.

\begin{figure}[t]
\centering
\begin{tikzpicture}
\node[anchor=south] at (1.0, 2.2) {\textbf{(a) Quality $\uparrow$}};

\begin{axis}[
    name=plot1,
    width=0.50\columnwidth,
    height=0.47\columnwidth,
    ylabel={\textbf{Dist-Sim}},
    xmin=0.5, xmax=3.5,
    ymin=0.60, ymax=1.05,
    xtick={1, 2, 3},
    xticklabels={},
    ytick={0.7, 0.8, 0.9, 1.0},
    tick label style={font=\scriptsize},
    label style={font=\small},
    grid=major,
]
\addplot[color=magenta!80!black, mark=pentagon*, thick, only marks] coordinates {(1, 1.00)};
\addplot[color=red, mark=square*, thick] coordinates {(1, 0.74) (2, 0.74) (3, 0.74)};
\addplot[color=orange, mark=triangle*, thick] coordinates {(1, 0.71) (2, 0.71) (3, 0.71)};
\addplot[color=purple, mark=diamond*, thick] coordinates {(1, 0.71) (2, 0.71) (3, 0.71)};
\addplot[color=blue, mark=*, thick, line width=1.2pt] coordinates {(1, 0.73) (2, 0.69) (3, 0.68)};
\end{axis}

\begin{axis}[
    at={(plot1.south)},
    anchor=north,
    yshift=-0.5cm,
    name=plot2,
    width=0.50\columnwidth,
    height=0.47\columnwidth,
    xlabel={Queries/Doc},
    ylabel={\textbf{Len-Sim}},
    xmin=0.5, xmax=3.5,
    ymin=0.40, ymax=1.05,
    xtick={1, 2, 3},
    ytick={0.5, 0.7, 0.9},
    tick label style={font=\scriptsize},
    label style={font=\small},
    grid=major,
]
\addplot[color=magenta!80!black, mark=pentagon*, thick, only marks] coordinates {(1, 1.00)};
\addplot[color=red, mark=square*, thick] coordinates {(1, 0.52) (2, 0.52) (3, 0.52)};
\addplot[color=orange, mark=triangle*, thick] coordinates {(1, 0.51) (2, 0.51) (3, 0.51)};
\addplot[color=purple, mark=diamond*, thick] coordinates {(1, 0.63) (2, 0.63) (3, 0.63)};
\addplot[color=blue, mark=*, thick, line width=1.2pt] coordinates {(1, 0.56) (2, 0.49) (3, 0.51)};
\end{axis}

\node[anchor=south] at (5.0, 2.2) {\textbf{(b) Diversity $\downarrow$}};

\begin{axis}[
    at={(plot1.east)},
    anchor=west,
    xshift=1.8cm,
    name=plot3,
    width=0.50\columnwidth,
    height=0.47\columnwidth,
    ylabel={\textbf{CE}},
    xmin=0.5, xmax=3.5,
    ymin=0, ymax=0.90,
    xtick={1, 2, 3},
    xticklabels={},
    ytick={0, 0.2, 0.4, 0.6, 0.8},
    tick label style={font=\scriptsize},
    label style={font=\small},
    grid=major,
]
\addplot[color=magenta!80!black, mark=pentagon*, thick, only marks] coordinates {(1, 0.75)};
\addplot[color=red, mark=square*, thick] coordinates {(1, 0.70) (2, 0.76) (3, 0.78)};
\addplot[color=orange, mark=triangle*, thick] coordinates {(1, 0.59) (2, 0.67) (3, 0.69)};
\addplot[color=purple, mark=diamond*, thick] coordinates {(1, 0.51) (2, 0.41) (3, 0.42)};
\addplot[color=blue, mark=*, thick, line width=1.2pt] coordinates {(1, 0.81) (2, 0.14) (3, 0.10)};
\end{axis}

\begin{axis}[
    at={(plot3.south)},
    anchor=north,
    yshift=-0.5cm,
    name=plot4,
    width=0.50\columnwidth,
    height=0.47\columnwidth,
    xlabel={Queries/Doc},
    ylabel={\textbf{Self-BLEU}},
    xmin=0.5, xmax=3.5,
    ymin=0, ymax=1.0,
    xtick={1, 2, 3},
    ytick={0, 0.2, 0.4, 0.6, 0.8},
    tick label style={font=\scriptsize},
    label style={font=\small},
    grid=major,
]
\addplot[color=magenta!80!black, mark=pentagon*, thick, only marks] coordinates {(1, 0.95)};
\addplot[color=red, mark=square*, thick] coordinates {(1, 0.59) (2, 0.65) (3, 0.76)};
\addplot[color=orange, mark=triangle*, thick] coordinates {(1, 0.54) (2, 0.59) (3, 0.71)};
\addplot[color=purple, mark=diamond*, thick] coordinates {(1, 0.54) (2, 0.33) (3, 0.45)};
\addplot[color=blue, mark=*, thick, line width=1.2pt] coordinates {(1, 0.79) (2, 0.14) (3, 0.16)};
\end{axis}

\node[anchor=base] at (0.0, -4.0) {\footnotesize\raisebox{1.5pt}{\textcolor{magenta!80!black}{\pgfuseplotmark{pentagon*}}} Sup.};
\node[anchor=base] at (1.3, -4.0) {\footnotesize\raisebox{0.5pt}{\textcolor{red}{$\blacksquare$}} InPars};
\node[anchor=base] at (2.6, -4.0) {\footnotesize\raisebox{0pt}{\textcolor{orange}{$\blacktriangle$}} SAP};
\node[anchor=base] at (3.9, -4.0) {\footnotesize\raisebox{0.5pt}{\textcolor{purple}{$\blacklozenge$}} DRAMA};
\node[anchor=base] at (5.2, -4.0) {\footnotesize\raisebox{-1pt}{\textcolor{blue}{$\bullet$}} Ours};
\end{tikzpicture}
\caption{Quality and Diversity metrics as the number of queries per document increases. \textbf{(a) Quality ($\uparrow$)}: Dist-Sim and Len-Sim measure similarity to human-annotated queries, where higher values indicate more human-like quality. \textbf{(b) Diversity ($\downarrow$)}: CE and Self-BLEU measure query similarity, where lower values indicate higher diversity. Few-shot methods become \textit{less diverse} (values increase), while our method becomes dramatically \textit{more diverse} (scores drop from $\sim$0.8 to $\sim$0.1). DRAGON-S is excluded as it uses unsupervised sentence cropping rather than LLM generation, making quality and diversity metrics not comparable. Full data in Appendix Table~\ref{tab:qd_full}.}
\label{fig:qd_tradeoff}
\end{figure}

\begin{table*}[!t]
\centering
\small
\resizebox{\textwidth}{!}{
\begin{tabular}{l|c|cccc|cccc}
\toprule
& & \multicolumn{4}{c|}{\textbf{Contriever}} & \multicolumn{4}{c}{\textbf{RetroMAE}} \\
\textbf{Method} & \textbf{Q/Doc} & \textbf{TREC-DL (2)} & \textbf{BEIR (14)} & \textbf{BRIGHT (12)} & \textbf{Multi-hop (4)} & \textbf{TREC-DL (2)} & \textbf{BEIR (14)} & \textbf{BRIGHT (12)} & \textbf{Multi-hop (4)} \\
\midrule
\textbf{Pretrained} & - & 41.68 & 28.73 & 3.84 & 39.70 & 11.52 & 14.27 & 5.20 & 18.45 \\
\textbf{Supervised} & 1 & \textbf{57.52} & 40.39 & 7.11 & 50.41 & \underline{56.26} & 38.47 & 6.33 & 51.15 \\
\midrule
\multicolumn{10}{c}{\textit{Non-LLM Data Augmentation}} \\
\midrule
\textbf{DRAGON-S} & 1 & 4.10 & 7.83 & 1.52 & 2.37 & 3.90 & 6.73 & 1.39 & 2.22 \\
DRAGON-S$^\dagger$ & 2 & 4.80 & 7.29 & 1.51 & 2.51 & 5.11 & 6.26 & 1.48 & 1.71 \\
DRAGON-S$^\dagger$ & 3 & 3.88 & 8.31 & 1.64 & 3.27 & 3.20 & 6.09 & 1.60 & 1.34 \\
\midrule
\multicolumn{10}{c}{\textit{Few-shot LLM Query Synthesis}} \\
\midrule
\textbf{InPars-GBQ} & 1 & \underline{56.33} & 41.31 & 7.84 & 52.72 & 56.07 & 39.02 & 6.98 & 52.39 \\
InPars-GBQ$^\dagger$ & 2 & 55.31 & 41.26 & 8.44 & 52.58 & \textbf{56.71} & 39.73 & 7.22 & 53.50 \\
InPars-GBQ$^\dagger$ & 3 & 54.35 & 41.15 & 8.22 & 53.26 & 56.08 & 39.89 & 7.82 & 53.79 \\
\midrule
\textbf{SAP} & 1 & 53.53 & 41.37 & 9.49 & 52.58 & 54.53 & 39.69 & 7.43 & 54.85 \\
SAP$^\dagger$ & 2 & 53.08 & \textbf{41.50} & 8.95 & 53.32 & 54.12 & \textbf{40.20} & \textbf{8.63} & 55.84 \\
SAP$^\dagger$ & 3 & 53.42 & 41.41 & 9.00 & 52.78 & 54.30 & \underline{40.04} & 7.94 & 53.96 \\
\midrule
\textbf{DRAMA} & 1 & 49.61 & 37.36 & 9.43 & 50.72 & 52.12 & 35.38 & 6.42 & 49.42 \\
DRAMA$^\dagger$ & 2 & 50.10 & 37.75 & \textbf{9.83} & 51.44 & 51.50 & 34.32 & 6.29 & 48.74 \\
DRAMA$^\dagger$ & 3 & 48.14 & 37.47 & 9.50 & 51.41 & 51.71 & 34.69 & 6.06 & 48.97 \\
\midrule
\multicolumn{10}{c}{\textit{Zero-shot LLM Query Synthesis (Ours)}} \\
\midrule
\textbf{Ours} & 1 & 52.94 & \underline{41.43} & 8.57 & 52.77 & 54.00 & 39.31 & 7.72 & 54.56 \\
\rowcolor{gray!20} \textbf{Ours} & 2 & 53.97 & 40.94 & 8.95 & 54.20 & 52.66 & 39.14 & \underline{8.07} & \textbf{58.50} \\
\rowcolor{gray!20} \textbf{Ours} & 3 & 53.34 & 40.82 & 9.12 & \textbf{55.22} & 53.89 & 38.61 & 7.69 & \underline{57.09} \\
\midrule
\multicolumn{10}{c}{\textit{Ours with 8k Documents (10\% of baselines)}} \\
\midrule
\textbf{Ours (8k)} & 2 & 50.33 & 40.72 & \underline{9.50} & 54.34 & 51.26 & 38.97 & 7.40 & 56.91 \\
\textbf{Ours (8k)} & 5 & 48.86 & 40.70 & 9.14 & \underline{54.40} & 49.37 & 38.85 & 8.02 & 56.35 \\
\bottomrule
\end{tabular}
}
\caption{Multi-query results (NDCG@10). \textbf{Pretrained}: original checkpoints. \textbf{Supervised}: fine-tuned with human-annotated queries. All methods use 80k documents except ``Ours (8k)''. Best in \textbf{bold}, second-best \underline{underlined}. $^\dagger$Extended with temperature=0.7.}
\label{tab:main_results}
\end{table*}

\begin{table*}[!htbp]
\centering
\small
\resizebox{\textwidth}{!}{
\begin{tabular}{l|c|ccccc|ccccc}
\toprule
& & \multicolumn{5}{c|}{\textbf{Contriever}} & \multicolumn{5}{c}{\textbf{RetroMAE}} \\
\textbf{Method} & \textbf{Q/Doc} & \textbf{Novel} & \textbf{Hotpot} & \textbf{MuSiQue} & \textbf{2Wiki} & \textbf{Avg} & \textbf{Novel} & \textbf{Hotpot} & \textbf{MuSiQue} & \textbf{2Wiki} & \textbf{Avg} \\
\midrule
\textbf{Pretrained} & - & 37.84 & 41.01 & 32.11 & 47.85 & 39.70 & 21.20 & 20.88 & 12.55 & 19.18 & 18.45 \\
\textbf{Supervised} & 1 & 54.53 & 52.01 & 33.72 & 61.39 & 50.41 & 58.61 & 48.92 & 33.42 & 63.65 & 51.15 \\
\midrule
\textbf{DRAGON-S} & 1 & 2.34 & 1.21 & 2.05 & 3.88 & 2.37 & 1.87 & 1.05 & 1.63 & 4.34 & 2.22 \\
DRAGON-S$^\dagger$ & 2 & 1.93 & 1.13 & 1.84 & 5.14 & 2.51 & 1.67 & 0.75 & 1.37 & 3.04 & 1.71 \\
DRAGON-S$^\dagger$ & 3 & 2.48 & 1.81 & 2.31 & 6.49 & 3.27 & 1.64 & 0.59 & 1.12 & 1.99 & 1.34 \\
\midrule
\textbf{InPars-GBQ} & 1 & 56.49 & 53.85 & 35.62 & 64.94 & 52.73 & 61.38 & 49.91 & 33.95 & 64.31 & 52.39 \\
InPars-GBQ$^\dagger$ & 2 & 55.77 & 53.82 & 35.89 & 64.83 & 52.58 & 61.63 & 51.95 & 34.25 & 66.19 & 53.50 \\
InPars-GBQ$^\dagger$ & 3 & 56.99 & 53.78 & 36.27 & 66.00 & 53.26 & 62.48 & 52.94 & 34.25 & 65.50 & 53.79 \\
\midrule
\textbf{SAP} & 1 & 57.41 & 55.64 & 35.74 & 61.52 & 52.58 & 66.09 & 53.87 & 35.51 & 63.93 & 54.85 \\
SAP$^\dagger$ & 2 & 58.91 & \underline{56.35} & 35.55 & 62.45 & 53.31 & 68.37 & 55.37 & 35.19 & 64.43 & 55.84 \\
SAP$^\dagger$ & 3 & 57.22 & 55.93 & 35.70 & 62.27 & 52.78 & 66.60 & 53.15 & 34.71 & 61.38 & 53.96 \\
\midrule
\textbf{DRAMA} & 1 & 50.12 & 53.61 & 35.86 & 63.30 & 50.72 & 49.34 & 51.36 & 33.42 & 63.55 & 49.42 \\
DRAMA$^\dagger$ & 2 & 53.45 & 53.42 & 35.51 & 63.36 & 51.44 & 48.48 & 50.40 & 32.99 & 63.08 & 48.74 \\
DRAMA$^\dagger$ & 3 & 53.35 & 53.96 & 35.40 & 62.94 & 51.41 & 49.20 & 50.37 & 33.20 & 63.10 & 48.97 \\
\midrule
\textbf{Ours} & 1 & 55.08 & 55.58 & 35.39 & 65.02 & 52.77 & 64.20 & 54.38 & 34.64 & 65.03 & 54.56 \\
\rowcolor{gray!20} \textbf{Ours} & 2 & 60.20 & 55.49 & \underline{36.60} & 64.50 & 54.20 & \textbf{72.81} & 55.42 & \textbf{38.60} & \underline{67.18} & \textbf{58.50} \\
\rowcolor{gray!20} \textbf{Ours} & 3 & \textbf{62.49} & 55.62 & 36.54 & \textbf{66.24} & \textbf{55.22} & \underline{72.25} & 53.29 & 37.38 & 65.42 & \underline{57.09} \\
\midrule
\multicolumn{12}{c}{\textit{Ours with 8k Documents (10\% of baselines)}} \\
\midrule
\textbf{Ours (8k)} & 2 & 58.88 & 55.40 & \textbf{37.05} & \underline{66.03} & 54.34 & 64.52 & \textbf{56.91} & \underline{38.24} & \textbf{67.98} & 56.91 \\
\textbf{Ours (8k)} & 5 & \underline{62.14} & \textbf{56.55} & 35.72 & 63.21 & \underline{54.41} & 68.99 & \underline{55.96} & 36.50 & 63.93 & 56.35 \\
\bottomrule
\end{tabular}
}
\caption{Detailed multi-hop retrieval results for multi-query experiments (NDCG@10). 2Wiki=2WikiMultihopQA, Hotpot=HotpotQA, Novel=NovelHopQA.}
\label{tab:multihop_full}
\end{table*}

\section{Results}

We compare our method against baselines, then analyze which tasks benefit most from diversity.

Figure~\ref{fig:qd_tradeoff} shows the quality-diversity trade-off: our method produces slightly lower quality but dramatically higher diversity, with CE dropping to 0.10--0.14 compared to baselines' 0.51--0.78. Table~\ref{tab:main_results} presents retrieval results:

\paragraph{Our method excels on reasoning tasks.} Few-shot baselines produce homogeneous queries with CE=0.51--0.78, while our method generates truly diverse queries with CE$<$0.15. Despite simpler prompting, we achieve the best multi-hop performance, i.e., 55.22 vs.\ 53.26 for InPars-GBQ, while maintaining competitive BEIR at 41.43. DRAGON-S fails catastrophically on Multi-hop with only 2.37, confirming that sentence cropping alone is insufficient; DRAGON's success comes from mixing data sources, not query diversity.

\paragraph{Multi-hop benefits most.} Table~\ref{tab:multihop_full} shows our method achieves best average across all multi-hop datasets, with largest gains on NovelHopQA at +5.5 over InPars-GBQ. Figure~\ref{fig:diversity_multihop} confirms diverse training consistently outperforms paraphrase training. NovelHopQA benefits most due to variable hop depths of 1--4 and long contexts as shown in Table~\ref{tab:multihop_difficulty}, while 2WikiMultihopQA and HotpotQA often admit single-hop shortcuts \citep{trivedi2022musique}.

\begin{figure}[!htbp]
\centering
\begin{tikzpicture}
\begin{axis}[
    name=contriever,
    width=0.58\columnwidth,
    height=0.55\columnwidth,
    xlabel={},
    ylabel={\textbf{Multi-hop NDCG@10}},
    title={\textbf{(a) Contriever}},
    title style={at={(0.4,1)}},
    xmin=0, xmax=0.95,
    x dir=reverse,
    ymin=48, ymax=59,
    xtick={0.2, 0.4, 0.6, 0.8},
    ytick={49, 51, 53, 55, 57, 59},
    tick label style={font=\footnotesize},
    label style={font=\small},
    title style={font=\small},
    grid=major,
]
\addplot[color=magenta!80!black, mark=pentagon*, thick, only marks, mark size=3.5pt] coordinates {(0.75, 50.41)};
\addplot[color=red, mark=square*, thick, only marks, mark size=3pt] coordinates {(0.70, 52.72)};
\addplot[color=orange, mark=triangle*, thick, only marks, mark size=3.5pt] coordinates {(0.59, 52.58)};
\addplot[color=purple, mark=diamond*, thick, only marks, mark size=3.5pt] coordinates {(0.51, 50.72)};
\addplot[color=blue, mark=*, thick, only marks, mark size=3.5pt] coordinates {(0.10, 55.22)};
\end{axis}

\begin{axis}[
    at={(contriever.east)},
    anchor=west,
    xshift=0.2cm,
    width=0.58\columnwidth,
    height=0.55\columnwidth,
    xlabel={},
    ylabel={},
    title={\textbf{(b) RetroMAE}},
    title style={at={(0.4,1)}},
    xmin=0, xmax=0.95,
    x dir=reverse,
    ymin=48, ymax=59,
    xtick={0.2, 0.4, 0.6, 0.8},
    ytick={49, 51, 53, 55, 57, 59},
    yticklabels={},
    tick label style={font=\footnotesize},
    label style={font=\small},
    title style={font=\small},
    grid=major,
]
\addplot[color=magenta!80!black, mark=pentagon*, thick, only marks, mark size=3.5pt] coordinates {(0.75, 51.15)};
\addplot[color=red, mark=square*, thick, only marks, mark size=3pt] coordinates {(0.70, 52.39)};
\addplot[color=orange, mark=triangle*, thick, only marks, mark size=3.5pt] coordinates {(0.59, 54.85)};
\addplot[color=purple, mark=diamond*, thick, only marks, mark size=3.5pt] coordinates {(0.51, 49.42)};
\addplot[color=blue, mark=*, thick, only marks, mark size=3.5pt] coordinates {(0.10, 58.50)};
\end{axis}

\node[anchor=north] at (2.8, -0.5) {\small CE (lower $=$ more diverse)};

\node[anchor=north] at (2.8, -1.2) {
\footnotesize
\raisebox{1.5pt}{\textcolor{magenta!80!black}{\pgfuseplotmark{pentagon*}}} Sup.\quad
\raisebox{0.5pt}{\textcolor{red}{$\blacksquare$}} InPars\quad
\raisebox{0pt}{\textcolor{orange}{$\blacktriangle$}} SAP\quad
\raisebox{0.5pt}{\textcolor{purple}{$\blacklozenge$}} DRAMA\quad
\raisebox{-1pt}{\textcolor{blue}{$\bullet$}} Ours
};
\end{tikzpicture}
\caption{Query diversity correlates with multi-hop retrieval performance. X-axis: CE (cross-encoder paraphrase ratio). Y-axis: NDCG@10 on Multi-hop benchmark. Our method with the lowest CE achieves the best performance on both retrievers.}
\label{fig:diversity_multihop}
\end{figure}

\begin{table}[!htbp]
\centering
\small
\begin{tabular*}{\columnwidth}{@{\extracolsep{\fill}}l|ccc@{}}
\toprule
\textbf{Dataset} & \textbf{H-M Gap} & \textbf{Hops} & \textbf{Context} \\
\midrule
NovelHopQA & -- & 1--4 & Long \\
HotpotQA & 9.6 & 2 & Short \\
MuSiQue & 28.2 & 2--4 & Short \\
2WikiMultihopQA & 3.7 & 2,4 & Short \\
\bottomrule
\end{tabular*}
\caption{Multi-hop dataset characteristics. H-M Gap: Human-Machine F1 gap \citep{trivedi2022musique}. NovelHopQA uniquely combines variable hops (1--4) with long context.}
\label{tab:multihop_difficulty}
\end{table}

\subsection{Controlled Diversity Experiments}
\label{sec:diversity_experiments}

We systematically vary diversity through: (1) \textbf{Prompting strategy}, comparing Paraphrase vs.\ Diverse at Q/Doc $\in$ \{5, 10, 20\}; (2) \textbf{Query quantity}, scaling Q/Doc from 1 to 20. This yields 14 conditions across 2 architectures.

\begin{table}[!htbp]
\centering
\small
\resizebox{\columnwidth}{!}{
\begin{tabular}{ll|c|cccc|c}
\toprule
\textbf{Model} & \textbf{Variant} & \textbf{Q/Doc} & \textbf{Novel} & \textbf{Hotpot} & \textbf{MuSiQue} & \textbf{2Wiki} & \textbf{Avg} \\
\midrule
\multirow{6}{*}{Contriever} & Paraphrase & 5 & 52.26 & 52.58 & 35.21 & \textbf{66.34} & 51.60 \\
& Diverse & 5 & \textbf{62.14} & \textbf{56.55} & \textbf{35.72} & 63.21 & \textbf{54.40} \\
\cmidrule{2-8}
& Paraphrase & 10 & 55.97 & 52.74 & \textbf{35.57} & \textbf{65.84} & 52.53 \\
& Diverse & 10 & \textbf{63.29} & \textbf{55.61} & 34.67 & 61.96 & \textbf{53.88} \\
\cmidrule{2-8}
& Paraphrase & 20 & 53.93 & 51.57 & \textbf{34.77} & \textbf{63.97} & 51.06 \\
& Diverse & 20 & \textbf{64.37} & \textbf{54.93} & 32.75 & 58.01 & \textbf{52.52} \\
\midrule
\multirow{6}{*}{RetroMAE} & Paraphrase & 5 & 59.33 & 50.25 & 33.36 & \textbf{64.96} & 51.97 \\
& Diverse & 5 & \textbf{68.99} & \textbf{55.96} & \textbf{36.50} & 63.93 & \textbf{56.35} \\
\cmidrule{2-8}
& Paraphrase & 10 & 59.64 & 51.51 & 32.89 & \textbf{64.06} & 52.02 \\
& Diverse & 10 & \textbf{71.17} & \textbf{54.50} & \textbf{35.57} & 62.40 & \textbf{55.91} \\
\cmidrule{2-8}
& Paraphrase & 20 & 60.73 & \textbf{50.82} & \textbf{32.83} & \textbf{63.87} & 52.06 \\
& Diverse & 20 & \textbf{69.91} & 50.61 & 32.78 & 56.64 & \textbf{52.49} \\
\bottomrule
\end{tabular}
}
\caption{NDCG@10 (\%) on multi-hop datasets comparing Paraphrase (CE$\approx$0.55) vs.\ Diverse (CE$\approx$0.04) training at different Q/Doc ratios on 8k documents. Diverse consistently outperforms on Novel (+9.88 to +10.44 for Contriever; +9.17 to +11.53 for RetroMAE) across all Q/Doc settings. 2Wiki consistently favors Paraphrase, while Hotpot and MuSiQue show mixed patterns. Full results in Appendix~\ref{sec:appendix_diversity_ablation}, Table~\ref{tab:diversity_ablation}.}
\label{tab:multihop_diversity}
\end{table}

\begin{table}[!htbp]
\centering
\small
\resizebox{\columnwidth}{!}{
\begin{tabular}{llcccc|c}
\toprule
\textbf{Variant} & \textbf{Q/Doc} & \textbf{Hotpot} & \textbf{2Wiki} & \textbf{MuSiQue} & \textbf{Novel} & \textbf{Avg} \\
\midrule
Paraphrase & 1 & 50.53 & 66.60 & 35.86 & \textbf{54.33} & \textbf{51.83} \\
Diverse & 1 & \textbf{51.92} & \textbf{67.99} & \textbf{37.03} & 50.30 & 51.81 \\
\midrule
Paraphrase & 5 & 52.58 & \textbf{66.34} & 35.21 & 52.26 & 51.60 \\
Diverse & 5 & \textbf{56.55} & 63.21 & \textbf{35.72} & \textbf{62.14} & \textbf{54.41} \\
\bottomrule
\end{tabular}}
\caption{Single-query ablation (Contriever). With Q/Doc=1, Diverse performs worse than Paraphrase on Novel. Improvement emerges with Q/Doc=5, confirming \textbf{multiple views} drive gains.}
\label{tab:single_query_ablation}
\end{table}

\paragraph{Diversity vs Paraphrase.} Table~\ref{tab:multihop_diversity} shows Diverse outperforms Paraphrase on NovelHopQA by +9.88/+9.66 and HotpotQA by +3.97/+5.71, while 2WikiMultihopQA slightly favors Paraphrase. This heterogeneity suggests diversity benefit varies with query complexity. Full results in Appendix~\ref{sec:appendix_diversity_ablation}.

\paragraph{Single-Query Ablation.} Does the improvement come from diverse query quality or from multiple queries? Table~\ref{tab:single_query_ablation} compares Q/Doc=1, i.e., single query per document. With single queries, Diverse performs slightly \textit{worse} than Paraphrase on NovelHopQA at 50.30 vs 54.33, as paraphrase queries follow supervised data patterns with higher individual quality. The improvement emerges only with Q/Doc=5 at 62.14, confirming that \textbf{multiple semantic views of the same document} drive the gains, rather than broader query distribution alone.

\subsection{Query Scaling Effects}

\begin{table}[!htbp]
\centering
\small
\resizebox{\columnwidth}{!}{
\begin{tabular}{l|c|cccc}
\toprule
\textbf{Model} & \textbf{Q/Doc} & \textbf{Novel} & \textbf{Hotpot} & \textbf{MuSiQue} & \textbf{2Wiki} \\
\midrule
\multirow{5}{*}{Contriever} & 1 & 50.67 & 53.33 & 37.01 & \textbf{68.32} \\
& 2 & 58.88 & 55.40 & \textbf{37.05} & 66.03 \\
& 5 & 62.14 & \textbf{56.55} & 35.72 & 63.21 \\
& 10 & 63.29 & 55.61 & 34.67 & 61.96 \\
& 20 & \textbf{64.37} & 54.93 & 32.75 & 58.01 \\
\midrule
\multirow{5}{*}{RetroMAE} & 1 & 56.11 & 52.10 & 33.12 & 64.90 \\
& 2 & 64.52 & \textbf{56.91} & \textbf{38.24} & \textbf{67.98} \\
& 5 & 68.99 & 55.96 & 36.50 & 63.93 \\
& 10 & \textbf{71.17} & 54.50 & 35.57 & 62.40 \\
& 20 & 69.91 & 50.61 & 32.78 & 56.64 \\
\bottomrule
\end{tabular}
}
\caption{NDCG@10 (\%) on four multi-hop datasets when scaling diverse queries per document (8k documents). Full results in Appendix~\ref{sec:appendix_query_scaling}, Table~\ref{tab:query_scaling}.}
\label{tab:multihop_scaling}
\end{table}

Table~\ref{tab:multihop_scaling} shows the effect of scaling Q/Doc from 1 to 20. Full results are in Appendix~\ref{sec:appendix_query_scaling}. Datasets show different optimal Q/Doc ratios: NovelHopQA benefits from more queries, peaking at Q/Doc=10--20, while 2WikiMultihopQA performs best with fewer queries at Q/Doc=1--2. This heterogeneity motivates our investigation of CDP in Section~\ref{sec:cdp}.

\section{The Complexity-Diversity Principle}
\label{sec:cdp}

The results reveal that diversity benefits vary substantially across datasets. We hypothesize that \textit{query complexity} is the key factor, and validate this through correlation analysis.

\subsection{CW Correlation Discovery}

Using CW (Eq.~\ref{eq:cw_method}) to approximate complexity, Table~\ref{tab:content_words} shows strong correlation between CW and diversity benefit ($r \geq 0.95$, 12/14 with $p<0.05$). NovelHopQA (CW=11.64) benefits most; 2WikiMultihopQA (CW=6.34) shows minimal benefit.

\begin{table}[t]
\centering
\small
\resizebox{\columnwidth}{!}{
\begin{tabular}{ll|cccc|cc}
\toprule
& & \multicolumn{4}{c|}{\textbf{$\Delta$NDCG@10 (\%)}} & \multicolumn{2}{c}{\textbf{Pearson}} \\
\textbf{Model} & \textbf{Comparison} & \textbf{Novel} & \textbf{Hotpot} & \textbf{MuSiQue} & \textbf{2Wiki} & $r$ & $p$ \\
\midrule
\multicolumn{2}{l|}{\textit{Content Words (CW)}} & \textit{11.64} & \textit{8.60} & \textit{8.64} & \textit{6.34} & & \\
\midrule
\multirow{8}{*}[-0.5em]{Contriever} & \multicolumn{7}{l}{\textit{Diverse $-$ Paraphrase}} \\
& Q/Doc=5 & +9.9 & +4.0 & +0.5 & $-$3.1 & 0.96 & 0.035* \\
& Q/Doc=10 & +7.3 & +2.9 & $-$0.9 & $-$3.9 & 0.95 & 0.054 \\
& Q/Doc=20 & +10.4 & +3.4 & $-$2.0 & $-$6.0 & 0.95 & 0.052 \\
\cmidrule{2-8}
& \multicolumn{7}{l}{\textit{Q/Doc=$M$ $-$ Q/Doc=1}} \\
& $M$=2 & +8.2 & +2.1 & +0.0 & $-$2.3 & 0.97 & 0.030* \\
& $M$=5 & +11.5 & +3.2 & $-$1.3 & $-$5.1 & 0.96 & 0.040* \\
& $M$=10 & +12.6 & +2.3 & $-$2.3 & $-$6.4 & 0.96 & 0.037* \\
& $M$=20 & +13.7 & +1.6 & $-$4.3 & $-$10.3 & 0.97 & 0.032* \\
\midrule
\multirow{8}{*}[-0.5em]{RetroMAE} & \multicolumn{7}{l}{\textit{Diverse $-$ Paraphrase}} \\
& Q/Doc=5 & +9.7 & +5.7 & +3.1 & $-$1.0 & 0.96 & 0.036* \\
& Q/Doc=10 & +11.5 & +3.0 & +2.7 & $-$1.7 & 0.99 & 0.008* \\
& Q/Doc=20 & +9.2 & $-$0.2 & $-$0.0 & $-$7.2 & 1.00 & <0.001* \\
\cmidrule{2-8}
& \multicolumn{7}{l}{\textit{Q/Doc=$M$ $-$ Q/Doc=1}} \\
& $M$=2 & +8.4 & +4.8 & +5.1 & +3.1 & 0.99 & 0.007* \\
& $M$=5 & +12.9 & +3.9 & +3.4 & $-$1.0 & 0.99 & 0.010* \\
& $M$=10 & +15.1 & +2.4 & +2.5 & $-$2.5 & 0.98 & 0.021* \\
& $M$=20 & +13.8 & $-$1.5 & $-$0.3 & $-$8.3 & 0.99 & 0.010* \\
\bottomrule
\end{tabular}
}
\caption{CW correlates with diversity benefit ($\Delta$NDCG@10). All 14 conditions show $r \geq 0.95$, with 12/14 significant (*$p<0.05$).}
\label{tab:content_words}
\end{table}

\subsection{Threshold Formalization}

Table~\ref{tab:cw_threshold} shows positive rates: NovelHopQA (CW=11.64) benefits in 100\% of conditions, while 2WikiMultihopQA (CW=6.34) in only 7\%. This suggests that high-complexity tasks, e.g., CW$>$10, consistently benefit from diversity-aware training.

\begin{table}[!htbp]
\centering
\small
\begin{tabular}{l|c|c}
\toprule
\textbf{CW Range} & \textbf{Positive Rate} & \textbf{Diversity} \\
\midrule
CW $<$ 7 & 7\% (1/14) & Avoid \\
CW 7--10 & 43--86\% & Test \\
CW $>$ 10 & 100\% (14/14) & Recommend \\
\bottomrule
\end{tabular}
\caption{CW-based diversity recommendations for multi-hop QA tasks. Positive rate indicates the proportion of 14 experimental conditions (7 Contriever + 7 RetroMAE) where diversity training improves performance ($\Delta$NDCG@10 $>$ 0). Linear regression across 56 data points (4 datasets $\times$ 14 conditions) yields threshold CW=7.9 (where $\Delta$=0). These recommendations apply only to OOD tasks where CW-diversity correlation holds ($r$=0.89, $p$$<$0.0001).}
\label{tab:cw_threshold}
\end{table}

We formalize our findings as the \textbf{Complexity-Diversity Principle (CDP)}: \textit{query complexity determines whether diversity helps or hurts}. Complex queries contain multiple content words that jointly define the information need, where no single word can capture it alone. Training on diverse formulations helps models learn to match based on the full semantic content rather than superficial word overlap. For simple queries dominated by few distinctive terms, diversity provides limited benefit: 2WikiMultihopQA (CW=6.34) shows negative $\Delta$NDCG@10 in 93\% of conditions.

\paragraph{Theoretical Interpretation.} Multi-query training acts as \textit{implicit regularization}: each query variant provides a different ``view'' of the document, similar to multi-view learning \citep{xu2013multiview}. Error analysis (Appendix~\ref{sec:appendix_error_analysis}) confirms this: M=3 corrects M=1's overfitting to superficial features. Across 4 benchmark types, 68\% of datasets benefit from Q/Doc$>$1 (Appendix~\ref{sec:appendix_qdoc}).

\subsection{CW-Weighted Training}

We validate CW-weighted training as described in Section~\ref{sec:methodology_cw} on supervised MS MARCO data. Table~\ref{tab:cw_weighting} shows CW-weighting improves OOD for Contriever/RetroMAE/Qwen3-Embedding: BEIR by +0.99/+2.83/+0.16 and BRIGHT by +0.53/+0.14/+0.50, with slight in-domain trade-off. Per-dataset breakdown in Appendix~\ref{sec:appendix_cw_full}. Unlike ReasonEmbed's RI \citep{reasonembed2025} which requires model inference on query-document pairs, CW can be computed from queries alone via simple word counting.

\begin{table}[!ht]
\centering
\small
\resizebox{\columnwidth}{!}{
\begin{tabular}{llcccc}
\toprule
\textbf{Model} & \textbf{Training} & \textbf{TREC-DL} & \textbf{BEIR} & \textbf{BRIGHT} & \textbf{Multi-hop} \\
\midrule
\multirow{2}{*}{Contriever} & Standard & \textbf{57.52} & 40.39 & 7.11 & 50.41 \\
& CW-w & 55.88 & \textbf{41.38} & \textbf{7.64} & \textbf{51.75} \\
\midrule
\multirow{2}{*}{RetroMAE} & Standard & \textbf{56.26} & 38.47 & 6.33 & \textbf{51.15} \\
& CW-w & 55.23 & \textbf{41.30} & \textbf{6.47} & 50.91 \\
\midrule
\multirow{2}{*}{Qwen3-Emb} & Standard & \textbf{54.72} & 42.22 & 10.22 & \textbf{49.18} \\
& CW-w & 53.96 & \textbf{42.38} & \textbf{10.72} & 48.59 \\
\bottomrule
\end{tabular}}
\caption{CW-weighted training on supervised MS MARCO (80k). CW-weighting improves OOD with slight in-domain trade-off.}
\label{tab:cw_weighting}
\end{table}

\subsection{Diversity as Data Substitute}

Can diversity substitute for document quantity? Figure~\ref{fig:cost_efficiency_multihop} shows that for high-CW tasks, NovelHopQA with CW=11.64 improves from 64.2 to 71.2 as documents decrease from 80k to 8k, demonstrating that diversity effectively substitutes for quantity, while 2WikiMQA with CW=6.34 consistently degrades. Practitioners can achieve strong performance on high-complexity tasks with only 10\% of documents. Full analysis in Appendix~\ref{sec:appendix_cost}.

\begin{figure}[!ht]
\centering
\begin{tikzpicture}
\begin{axis}[
    name=plot1,
    width=0.56\columnwidth,
    height=0.55\columnwidth,
    title={\textbf{Contriever}},
    title style={font=\small},
    xlabel={\#Docs (k)},
    ylabel={\textbf{NDCG@10}},
    xmode=log,
    log basis x=2,
    xmin=3, xmax=100,
    x dir=reverse,
    ymin=30, ymax=68,
    xtick={4, 8, 16, 40, 80},
    xticklabels={4, 8, 16, 40, 80},
    ytick={35, 45, 55, 65},
    tick label style={font=\footnotesize},
    label style={font=\small},
    grid=major,
]
\addplot[color=red!80!black, mark=square*, thick, line width=1pt] coordinates {
    (80, 65.02) (40, 63.74) (16, 60.24) (8, 61.96) (4, 56.68)
};
\addplot[color=orange, mark=triangle*, thick, line width=1pt] coordinates {
    (80, 55.58) (40, 55.45) (16, 55.12) (8, 55.61) (4, 53.94)
};
\addplot[color=teal, mark=diamond*, thick, line width=1pt] coordinates {
    (80, 35.39) (40, 36.21) (16, 33.85) (8, 34.67) (4, 32.35)
};
\addplot[color=blue!80!black, mark=*, thick, line width=1pt] coordinates {
    (80, 55.08) (40, 60.48) (16, 60.00) (8, 63.29) (4, 62.99)
};
\end{axis}

\begin{axis}[
    at={(plot1.east)},
    anchor=west,
    xshift=0.8cm,
    name=plot2,
    width=0.56\columnwidth,
    height=0.55\columnwidth,
    title={\textbf{RetroMAE}},
    title style={font=\small},
    xlabel={\#Docs (k)},
    xmode=log,
    log basis x=2,
    xmin=3, xmax=100,
    x dir=reverse,
    ymin=30, ymax=75,
    xtick={4, 8, 16, 40, 80},
    xticklabels={4, 8, 16, 40, 80},
    ytick={35, 45, 55, 65},
    tick label style={font=\footnotesize},
    label style={font=\small},
    grid=major,
]
\addplot[color=red!80!black, mark=square*, thick, line width=1pt] coordinates {
    (80, 65.03) (40, 68.09) (16, 60.87) (8, 62.40) (4, 55.88)
};
\addplot[color=orange, mark=triangle*, thick, line width=1pt] coordinates {
    (80, 54.38) (40, 53.01) (16, 50.81) (8, 51.50) (4, 48.90)
};
\addplot[color=teal, mark=diamond*, thick, line width=1pt] coordinates {
    (80, 34.64) (40, 37.87) (16, 34.87) (8, 35.57) (4, 32.50)
};
\addplot[color=blue!80!black, mark=*, thick, line width=1pt] coordinates {
    (80, 64.20) (40, 70.18) (16, 68.37) (8, 71.17) (4, 68.23)
};
\end{axis}

\path (plot1.south) -- (plot2.south) coordinate[midway] (legendpos);
\node[anchor=north, yshift=-1.1cm] at (legendpos) {
    \footnotesize
    \begin{tabular}{@{}c@{\quad}c@{\quad}c@{\quad}c@{}}
    \raisebox{-1pt}{\textcolor{blue!80!black}{$\bullet$}} Novel &
    \raisebox{0pt}{\textcolor{orange}{$\blacktriangle$}} Hotpot &
    \raisebox{0.5pt}{\textcolor{teal}{$\blacklozenge$}} MuSiQue &
    \raisebox{0.5pt}{\textcolor{red!80!black}{$\blacksquare$}} 2Wiki \\[-2pt]
    \scriptsize(CW\,11.6) & \scriptsize(CW\,8.6) & \scriptsize(CW\,8.6) & \scriptsize(CW\,6.3)
    \end{tabular}
};
\end{tikzpicture}
\caption{Document quantity vs.\ diversity trade-off. High-CW (NovelHopQA) improves as documents decrease; low-CW (2WikiMQA) degrades. High-CW tasks benefit from diversity over coverage.}
\label{fig:cost_efficiency_multihop}
\end{figure}

\section{Generalization}
\label{sec:validation}

We validate CDP across architectures, pipelines, and languages.

\subsection{Cross-Architecture Generalization}

\paragraph{Open-Source Generator.} Table~\ref{tab:mistral_generator} shows Mistral-7B-Instruct validates CDP: Diverse outperforms Paraphrase on all multi-hop datasets (+1.49 avg). GPT-4o-mini shows larger gains (+2.80), suggesting stronger LLMs generate more semantically distinct queries that better facilitate learning query diversity.

\begin{table}[!ht]
\centering
\small
\resizebox{\columnwidth}{!}{
\begin{tabular}{llcccc}
\toprule
\textbf{Generator} & \textbf{Variant} & \textbf{Hotpot} & \textbf{2Wiki} & \textbf{MuSiQue} & \textbf{Novel} \\
\midrule
Mistral-7B & Paraphrase & 54.81 & 63.82 & 34.96 & 60.91 \\
Mistral-7B & Diverse & \textbf{56.00} & \textbf{66.02} & \textbf{36.78} & \textbf{61.64} \\
\bottomrule
\end{tabular}}
\caption{Multi-hop NDCG@10 with Mistral-7B-Instruct-v0.3 as query generator (8k documents, Q/Doc=5, Contriever). CDP holds across generators: Diverse outperforms Paraphrase on all datasets.}
\label{tab:mistral_generator}
\end{table}

\begin{table}[!ht]
\centering
\small
\resizebox{\columnwidth}{!}{
\begin{tabular}{llcccc}
\toprule
\textbf{Retriever} & \textbf{Variant} & \textbf{Hotpot} & \textbf{2Wiki} & \textbf{MuSiQue} & \textbf{Novel} \\
\midrule
Qwen3-Emb & Paraphrase & 41.56 & 60.12 & 29.80 & 64.78 \\
Qwen3-Emb & Diverse & \textbf{50.40} & \textbf{62.06} & \textbf{31.50} & \textbf{74.68} \\
\bottomrule
\end{tabular}}
\caption{Multi-hop NDCG@10 with Qwen3-Embedding-0.6B as retriever (8k documents, Q/Doc=5, GPT-4o-mini generator).}
\label{tab:qwen_retriever}
\end{table}

\begin{table}[!htbp]
\centering
\small
\begin{tabular}{lcc}
\toprule
\textbf{Retriever} & \textbf{$\Delta$ Multi-hop (4)} & \textbf{$\Delta$ Novel} \\
\midrule
Contriever (BERT) & +2.80 & +9.88 \\
RetroMAE (BERT) & +4.38 & +9.66 \\
Qwen3-Emb (LLM) & \textbf{+5.59} & \textbf{+9.90} \\
\bottomrule
\end{tabular}
\caption{Diversity benefit (Diverse $-$ Paraphrase, 8k documents, Q/Doc=5) across retriever architectures.}
\label{tab:retriever_comparison}
\end{table}

\paragraph{LLM-based Retriever.} Tables~\ref{tab:qwen_retriever} and \ref{tab:retriever_comparison} show LLM-based Qwen3-Embedding-0.6B has \textit{larger} diversity benefits (+5.59) than BERT-based Contriever (+2.80) and RetroMAE (+4.38), suggesting modern retrievers can better leverage query diversity and making CDP increasingly relevant.

\subsection{Reasoning-Intensive Pipeline}

ReasonEmbed \citep{reasonembed2025} is a state-of-the-art reasoning-intensive retrieval pipeline. It synthesizes 82K training samples from BRIGHT via ReMixer and applies RI-weighted training via Redapter. The Reasoning Index (RI) measures reasoning intensity using the original query $q$, reasoning-augmented query $q'$, and document $D$:
\begin{equation}
\text{RI}_\theta(s) = \min\left(\frac{\mathcal{L}_{q,D}}{\mathcal{L}_{q',D}}, \kappa\right),
\label{eq:ri}
\end{equation}
where $\mathcal{L}$ is InfoNCE loss, $q$ is the original query, $q'$ is the reasoning-augmented query, and $\kappa$=5.0. Higher RI indicates stronger reasoning requirements. RI weights samples during training via batch normalization.

\paragraph{Experimental Setup.} We sample 8K documents from public ReasonEmbed data derived from BRIGHT and train Qwen3-Embedding-0.6B. We use original queries combined with ReMixer-generated reasoning queries as multi-query data. We use BRIGHT for in-domain and Multi-hop for OOD evaluation.

\paragraph{Results.} Table~\ref{tab:reasonembed} validates CDP on ReasonEmbed. Per-dataset breakdown is in Appendix~\ref{sec:appendix_reasonembed_full}. (1) \textbf{Multi-query}: multi-query outperforms query-only by +1.80 OOD, and multi-query+CW outperforms query-only+CW by +3.46 OOD. (2) \textbf{CW-weighting}: query-only+CW outperforms query-only by +1.00 OOD, and multi-query+CW outperforms multi-query by +2.66 OOD, with potential in-domain trade-off. (3) \textbf{Ours vs ReasonEmbed}: multi-query+RI at 33.20 surpasses ReasonEmbed baseline query-only+RI at 29.84 by +3.36 OOD. This confirms CDP transfers to state-of-the-art reasoning pipelines.

\begin{table}[t]
\centering
\small
\resizebox{\columnwidth}{!}{
\begin{tabular}{llcc}
\toprule
 & & \textbf{BRIGHT (12)} & \textbf{Multi-hop (4)} \\
\textbf{Method} & \textbf{Weight} & \textit{(in-domain)} & \textit{(OOD)} \\
\midrule
\multicolumn{4}{l}{\textit{w/ reasoning query}} \\
query-only (ReasonEmbed) & RI & \textbf{17.70} & 29.84 \\
query-only & RI$\times$CW & \underline{17.06} & 32.15 \\
reasoning-only & -- & 12.29 & 21.25 \\
reasoning-only & RI & 14.56 & 31.61 \\
reasoning-only & CW & 12.05 & 20.48 \\
reasoning-only & RI$\times$CW & 13.92 & 28.74 \\
multi-query & -- & 16.39 & 29.56 \\
multi-query & CW & 15.55 & 32.22 \\
multi-query & RI & 16.88 & \textbf{33.20} \\
multi-query & RI$\times$CW & 16.17 & \underline{32.64} \\
\midrule
\multicolumn{4}{l}{\textit{w/o reasoning query}} \\
query-only & -- & 15.95 & 27.76 \\
query-only & CW & \textbf{16.31} & \textbf{28.76} \\
\bottomrule
\end{tabular}}
\caption{Validation on ReasonEmbed data (8k documents) with Qwen3-Embedding-0.6B. Multi-query and CW-weighting both improve OOD, with potential in-domain trade-off. See Table~\ref{tab:reasonembed_full} in Appendix for per-dataset breakdown.}
\label{tab:reasonembed}
\end{table}

\subsection{Multilingual Validation}
\label{sec:multilingual}

\begin{table}[t]
\centering
\resizebox{\columnwidth}{!}{%
\begin{tabular}{lccccccc}
\toprule
& & \multicolumn{2}{c}{\textbf{In-Domain}} & \multicolumn{2}{c}{\textbf{Avg OOD}} & \\
\cmidrule(lr){3-4} \cmidrule(lr){5-6}
\textbf{Train} & \textbf{CW} & Standard & CW-w & Standard & CW-w & \textbf{$\Delta$OOD} \\
\midrule
ko & 1--25 & .635 & \textbf{.646} & .565 & \textbf{.627} & +10.9\% \\
ja & 1--11 & .739 & \textbf{.753} & .634 & \textbf{.662} & +4.5\% \\
ru & 1--19 & \textbf{.733} & .727 & .683 & \textbf{.690} & +1.2\% \\
ar & 1--12 & .698 & \textbf{.701} & \textbf{.677} & .672 & $-$0.7\% \\
zh & 1--9 & \textbf{.692} & .686 & \textbf{.551} & .500 & $-$9.3\% \\
fr & 1--7 & \textbf{.658} & .492 & \textbf{.601} & .249 & $-$58.6\% \\
\midrule
mix & --- & .746 & \textbf{.749} & --- & --- & --- \\
\bottomrule
\end{tabular}%
}
\caption{MIRACL cross-lingual evaluation. CW-w: CW-weighted training. \textbf{CW}: Content Word range. \textbf{$\Delta$OOD}: relative change in Avg OOD. Languages with broader CW distributions (ko, ja) show OOD gains; narrow distributions (fr, zh) degrade, possibly related to stopword configuration (Table~\ref{tab:stopword_issue}).}
\label{tab:miracl_results}
\end{table}

We validate CW-weighted training on MIRACL \citep{zhang2023miracl}, a multilingual benchmark covering 6 languages. Table~\ref{tab:miracl_results} shows effectiveness correlates with CW distribution width: Korean with CW 1--25 shows +10.9\% OOD improvement; Japanese with CW 1--11 shows +4.5\%; French with CW 1--7 degrades by $-$58.6\%, possibly due to stopword lists including question words. CDP generalizes across languages with diverse CW distributions, but may require language-specific stopword configuration. Details in Appendix~\ref{sec:appendix_multilingual_cw}.

\FloatBarrier
\section{Conclusion}

We discover a quality-diversity trade-off in query synthesis: quality benefits in-domain while diversity benefits OOD. Through the \textbf{Complexity-Diversity Principle (CDP)}, we show that query complexity, approximated by the simple CW metric, determines optimal diversity ($r$$\geq$0.95). Based on CDP, both multi-query synthesis and CW-weighted training improve OOD for high-complexity tasks, with compounded gains when combined.

\section*{Limitations}

We validate CDP on three dense retriever architectures; extending to sparse retrievers and late-interaction models would strengthen generalizability. CDP is validated primarily on reasoning-intensive tasks (multi-hop QA, BRIGHT); its predictive power for other task types requires further investigation.

CW is not a perfect measure of query complexity: it correlates with query length by design, and ignores individual word difficulty, syntactic structure, and semantic relationships. On datasets with uniformly high CW (e.g., BRIGHT: CW 18--134 vs.\ MS MARCO: CW 1--10), CW's discriminative power within training batches may decrease. However, our experiments show that even this simple proxy effectively guides training decisions. Future work could refine CW (e.g., better stopword curation) or explore more sophisticated measures such as model-based complexity estimation. Our multilingual evaluation (Section~\ref{sec:multilingual}) shows CDP generalizes across languages with sufficiently diverse CW distributions, but performance degrades for languages with narrow CW ranges; we hypothesize this may be related to stopword configuration, though further investigation is needed.

\section*{Ethics Statement}

We use publicly available datasets and both commercial and open-source LLMs. No personal data is collected. Synthetic queries may inherit LLM biases.

\bibliography{custom}

\appendix
\floatplacement{figure}{!htbp}
\floatplacement{table}{!htbp}

\section{Algorithm}
\label{sec:appendix_algorithm}

Algorithm~\ref{alg:method} provides the complete pseudocode for our zero-shot multi-query generation approach, including the prompt tuning phase where Q-D metrics guide prompt selection.

\begin{algorithm}[htbp]
\caption{Zero-shot Multi-Query Generation with Q-D Guided Prompt Tuning}
\label{alg:method}
\begin{algorithmic}[1]
\REQUIRE Corpus $\mathcal{D}$, queries per doc $M$, LLM $\mathcal{L}$, target task (in-domain/OOD)
\ENSURE Training set $\mathcal{T}$

\STATE \textbf{// Phase 1: Prompt Tuning with Q-D Metrics}
\STATE $\mathcal{D}_{\text{sample}} \gets$ sample $n$ documents from $\mathcal{D}$ \COMMENT{e.g., $n$=100}
\STATE $\mathcal{P} \gets \{p_{\text{paraphrase}}, p_{\text{diverse}}, ...\}$ \COMMENT{Candidate prompts}
\FOR{each prompt $p \in \mathcal{P}$}
    \STATE Generate queries $\mathcal{Q}_p$ using $p$ on $\mathcal{D}_{\text{sample}}$
    \STATE Compute Q-D metrics: CE$_p$, Self-BLEU$_p$
\ENDFOR
\STATE Select $p^* \gets$ prompt matching target diversity \COMMENT{See Fig.~\ref{fig:method_overview}}
\STATE \hspace{1em} In-domain: CE $>$ 0.5, Self-BLEU $>$ 0.5
\STATE \hspace{1em} OOD: CE $<$ 0.5, Self-BLEU $<$ 0.5

\STATE \textbf{// Phase 2: Full-scale Query Generation}
\STATE $\mathcal{T} \gets \emptyset$
\FOR{each document $d \in \mathcal{D}$}
    \STATE $\{q_1, ..., q_M\} \gets \mathcal{L}(p^*(d, M), \text{temp}=0)$ \COMMENT{Single call}
    \FOR{$i = 1$ to $M$}
        \STATE $\mathcal{T} \gets \mathcal{T} \cup \{(q_i, d)\}$
    \ENDFOR
\ENDFOR
\STATE Train retriever $f_\theta$ on $\mathcal{T}$ with InfoNCE loss
\end{algorithmic}
\end{algorithm}

\section{Query Generation Examples}
\label{sec:appendix_query_examples}

Table~\ref{tab:paraphrase_queries} shows the 20 queries generated by the zero-shot paraphrase method for the same document used in Table~\ref{tab:fewshot_vs_zeroshot}. Unlike our diverse method, paraphrase queries all follow the same question pattern with only surface-level variations.

\begin{table*}[!htbp]
\centering
\scriptsize
\begin{tcolorbox}[
  width=0.98\textwidth,
  colback=white,
  colframe=gray!40,
  colbacktitle=gray!15,
  coltitle=black,
  title=\textbf{Document:} \textit{Results-Based Accountability (RBA) is a disciplined way of thinking and taking action that communities can use to improve the lives of children, youth, families, adults and the community as a whole. RBA is also used by organizations to improve the performance of their programs...},
  fonttitle=\scriptsize,
  boxrule=0.5pt,
  arc=4pt,
  left=4pt, right=4pt, top=4pt, bottom=4pt
]

\renewcommand{\arraystretch}{1.1}
\begin{tabular}{p{0.46\textwidth}p{0.46\textwidth}}
\multicolumn{2}{c}{\cellcolor{orange!15}\textbf{Zero-shot Paraphrase (20 queries)}} \\
\midrule
1. What is RBA and how does it benefit communities? & 11. How does RBA facilitate collective action among leaders? \\
2. How can communities utilize RBA to enhance lives? & 12. What is the significance of community well-being in RBA? \\
3. In what ways does RBA contribute to community improvement? & 13. How does RBA measure community success? \\
4. What are the main advantages of implementing RBA? & 14. What strategies can communities adopt using RBA? \\
5. How does RBA help organizations improve effectiveness? & 15. How does RBA support safe and clean neighborhoods? \\
6. What role does community impact play in RBA? & 16. What is the impact of RBA on program performance? \\
7. How does RBA address the needs of children and families? & 17. How can RBA create better job opportunities? \\
8. What is the purpose of using RBA in community initiatives? & 18. What is the relationship between RBA and leadership? \\
9. How can leaders in a community apply RBA? & 19. How does RBA influence readiness of children for school? \\
10. What outcomes can be expected from implementing RBA? & 20. What are the key components of community impact in RBA? \\
\multicolumn{2}{l}{\textcolor{gray}{$\hookrightarrow$ All queries follow the same ``What/How does RBA...'' pattern with surface-level variations only}} \\
\end{tabular}
\end{tcolorbox}
\caption{Queries generated by the zero-shot paraphrase method for the first document in MS MARCO. Unlike our diverse method (Table~\ref{tab:fewshot_vs_zeroshot}), all queries follow similar question patterns, lacking format diversity (no keywords, statements, or comparisons).}
\label{tab:paraphrase_queries}
\end{table*}

\section{Implementation Details}
\label{sec:appendix_implementation}

\paragraph{Training Configuration}
We train all retriever models using the sentence-transformers library. We use AdamW optimizer with $\beta_1=0.9$, $\beta_2=0.98$, $\epsilon=10^{-8}$, and weight decay of 0.01. The learning rate is set to $2 \times 10^{-6}$ for Contriever and $5 \times 10^{-6}$ for RetroMAE, selected via grid search using supervised human-annotated data on the MS MARCO development set, then applied uniformly to all methods. We use cosine learning rate decay without warmup. The batch size is 128, and models are trained for up to 100 epochs with gradient clipping (max norm 1.0). We use InfoNCE loss \citep{oord2018infonce} with scale factor 20.0 as the loss function, and cosine similarity for scoring. Training employs FP16 mixed precision with gradient checkpointing enabled. Checkpoints are saved and evaluated every 500 steps, with model selection based on NDCG@10 on the MS MARCO development set. We use the official Contriever (\texttt{facebook/contriever}) and RetroMAE (\texttt{Shitao/RetroMAE}) checkpoints as initialization.

\paragraph{Temperature Setting for LLM-based Query Generation}
For our zero-shot method, we use temperature=0 to ensure reproducibility, as our prompt explicitly instructs the LLM to generate multiple diverse queries in a single API call. For few-shot baselines (InPars-GBQ, SAP, DRAMA), we use temperature=0.7 when generating multiple queries per document, since few-shot prompts typically produce a single query per call and require multiple sampling to obtain M queries.

\paragraph{Learning Rate Selection}
We conduct a learning rate sweep using supervised human-annotated MS MARCO data, selecting from $\{2 \times 10^{-7}, 5 \times 10^{-7}, 1 \times 10^{-6}, 2 \times 10^{-6}, 5 \times 10^{-6}, 1 \times 10^{-5}\}$. Table~\ref{tab:lr_sensitivity} shows the best learning rate at different training checkpoints. The selected learning rate is then applied uniformly to all data synthesis methods.

\begin{table}[htbp]
\centering
\small
\begin{tabular}{l|l|cc}
\toprule
\textbf{Retriever} & \textbf{Checkpoint} & \textbf{Best LR} & \textbf{NDCG@10} \\
\midrule
\multirow{4}{*}{Contriever} & 10k steps & $1 \times 10^{-5}$ & 92.86 \\
& 20k steps & $2 \times 10^{-6}$ & 92.83 \\
& 30k steps & $2 \times 10^{-6}$ & 92.87 \\
& \cellcolor{gray!15}Overall best & \cellcolor{gray!15}$2 \times 10^{-6}$ & \cellcolor{gray!15}\textbf{92.87} \\
\midrule
\multirow{4}{*}{RetroMAE} & 10k steps & $1 \times 10^{-5}$ & 91.52 \\
& 20k steps & $5 \times 10^{-6}$ & 91.87 \\
& 30k steps & $5 \times 10^{-6}$ & 91.73 \\
& \cellcolor{gray!15}Overall best & \cellcolor{gray!15}$5 \times 10^{-6}$ & \cellcolor{gray!15}\textbf{91.87} \\
\midrule
\multirow{4}{*}{Qwen3-Emb} & 1k steps & $5 \times 10^{-6}$ & 84.50 \\
& 5k steps & $5 \times 10^{-6}$ & 86.53 \\
& 10k steps & $5 \times 10^{-6}$ & 86.31 \\
& \cellcolor{gray!15}Overall best & \cellcolor{gray!15}$5 \times 10^{-6}$ & \cellcolor{gray!15}\textbf{86.53} \\
\bottomrule
\end{tabular}
\caption{Learning rate sensitivity analysis on the MS MARCO development set. We report the best learning rate at different training checkpoints and the overall best throughout training. Contriever and RetroMAE are trained for 100 epochs (checkpoints at 10k, 20k, 30k steps), while Qwen3-Embedding is trained for 20 epochs (checkpoints at 1k, 5k, 10k steps). The overall best learning rates ($2 \times 10^{-6}$ for Contriever, $5 \times 10^{-6}$ for RetroMAE and Qwen3-Embedding) are used for all experiments.}
\label{tab:lr_sensitivity}
\end{table}

\section{Full Quality-Diversity Metrics}
\label{sec:appendix_qd}

We define four Q-D metrics. For \textit{quality} (higher = better): (1) \textbf{Dist-Sim}, cosine similarity between synthetic ($q^s$) and human ($q^h$) query embeddings using BGE-M3 \citep{chen2024bge}; (2) \textbf{Len-Sim}, normalized length similarity. For \textit{diversity} (lower = more diverse): (1) \textbf{CE}, the proportion of query pairs with cross-encoder score $>$0.5; (2) \textbf{Self-BLEU}, average BLEU-4 score among queries:
\begin{align}
\text{Dist-Sim} &= \textstyle\frac{1}{N}\sum_{i} \cos(\mathbf{e}_{q^s_i}, \mathbf{e}_{q^h_i}) \\
\text{Len-Sim} &= 1 - |l_s - l_h| / \max(l_s, l_h) \\
\text{CE} &= \textstyle\frac{1}{|\mathcal{P}|}\sum_{i<j} \mathbb{1}[s(q_i, q_j) > 0.5] \\
\text{Self-BLEU} &= \textstyle\frac{1}{M}\sum_{i} \text{BLEU-4}(q_i, \{q_j\}_{j \neq i})
\end{align}

Table~\ref{tab:qd_full} provides the complete metrics for all methods and configurations.

\begin{table}[htbp]
\centering
\small
\resizebox{\columnwidth}{!}{
\begin{tabular}{l|c|cc|cc}
\toprule
\multirow{2}{*}{\textbf{Method}} & \multirow{2}{*}{\textbf{Q/Doc}} & \multicolumn{2}{c|}{\textbf{Quality}} & \multicolumn{2}{c}{\textbf{Diversity}} \\
& & Dist-Sim$\uparrow$ & Len-Sim$\uparrow$ & CE$\downarrow$ & Self-BLEU$\downarrow$ \\
\midrule
Supervised & 1 & 1.00 & 1.00 & 0.75 & 0.95 \\
\midrule
\multirow{3}{*}{InPars-GBQ} & 1 & 0.74 & 0.52 & 0.70 & 0.59 \\
& 2$^\dagger$ & 0.74 & 0.52 & 0.76 & 0.65 \\
& 3$^\dagger$ & 0.74 & 0.52 & 0.78 & 0.76 \\
\midrule
\multirow{3}{*}{SAP} & 1 & 0.71 & 0.51 & 0.59 & 0.54 \\
& 2$^\dagger$ & 0.71 & 0.51 & 0.67 & 0.59 \\
& 3$^\dagger$ & 0.71 & 0.51 & 0.69 & 0.71 \\
\midrule
\multirow{3}{*}{DRAMA} & 1 & 0.71 & 0.63 & 0.51 & 0.54 \\
& 2$^\dagger$ & 0.71 & 0.63 & 0.41 & 0.33 \\
& 3$^\dagger$ & 0.71 & 0.63 & 0.42 & 0.45 \\
\midrule
Ours & 1 & 0.73 & 0.56 & 0.81 & 0.79 \\
\rowcolor{gray!20} Ours & 2 & 0.69 & 0.49 & 0.14 & 0.14 \\
\rowcolor{gray!20} Ours & 3 & 0.68 & 0.51 & 0.10 & 0.16 \\
\bottomrule
\end{tabular}
}
\caption{Full quality and diversity metrics. Quality metrics (Dist-Sim, Len-Sim) measure similarity to human-written queries. Diversity metrics (CE, Self-BLEU) measure variation among generated queries for the same document. Our method achieves dramatically lower CE and Self-BLEU at Q/Doc$>$1, indicating truly diverse query generation. $^\dagger$Extended by us for fair comparison; original methods generate only one query per document.}
\label{tab:qd_full}
\end{table}

\section{Dataset Statistics}
\label{sec:appendix_datasets}

Table~\ref{tab:dataset_stats} provides detailed statistics for all evaluation datasets described in Section~\ref{sec:experiments}.

\begin{table*}[htbp]
\centering
\resizebox{\textwidth}{!}{%
\begin{tabular}{l|l|rrrrrrr}
\toprule
\textbf{Type} & \textbf{Dataset} & \textbf{\#Queries} & \textbf{\#Docs} & \textbf{Query Len} & \textbf{Doc Len} & \textbf{Docs/Query} & \textbf{Queries/Doc} & \textbf{CW} \\
\midrule
\multirow{2}{*}{\textbf{TREC-DL}} & trec-dl-2019 & 43 & 8,841,823 & 33 & 335 & 215 & 1 & 3.14 \\
& trec-dl-2020 & 54 & 8,841,823 & 34 & 335 & 211 & 1 & 3.56 \\
\midrule
\multirow{14}{*}{\textbf{BEIR}} & arguana & 1,406 & 8,674 & 1,193 & 1,030 & 1 & 1 & 81.55 \\
& climate-fever & 1,535 & 5,416,593 & 123 & 539 & 3 & 3 & 11.36 \\
& cqadupstack & 13,145 & 457,199 & 50 & 932 & 2 & 1 & 4.88 \\
& dbpedia-entity & 400 & 4,635,922 & 34 & 310 & 109 & 1 & 3.74 \\
& fever & 6,666 & 5,416,568 & 50 & 539 & 1 & 5 & 5.11 \\
& fiqa & 648 & 57,638 & 63 & 767 & 3 & 1 & 6.08 \\
& hotpotqa & 7,405 & 5,233,329 & 92 & 289 & 2 & 1 & 8.60 \\
& nfcorpus & 323 & 3,633 & 22 & 1,591 & 38 & 4 & 2.55 \\
& nq & 3,452 & 2,681,468 & 48 & 493 & 1 & 1 & 4.59 \\
& quora & 10,000 & 522,931 & 52 & 62 & 2 & 1 & 4.49 \\
& scidocs & 1,000 & 25,657 & 72 & 1,204 & 30 & 1 & 7.61 \\
& scifact & 300 & 5,183 & 90 & 1,499 & 1 & 1 & 8.36 \\
& trec-covid & 50 & 171,332 & 69 & 1,118 & 1,327 & 2 & 5.72 \\
& webis-touche2020 & 49 & 382,545 & 43 & 1,720 & 45 & 1 & 4.06 \\
\midrule
\multirow{12}{*}{\textbf{BRIGHT}} & aops & 111 & 188,002 & 320 & 754 & 5 & 5 & 16.74 \\
& biology & 103 & 57,359 & 523 & 330 & 4 & 1 & 34.61 \\
& earth\_science & 116 & 121,249 & 477 & 338 & 5 & 1 & 33.39 \\
& economics & 103 & 50,220 & 740 & 395 & 8 & 1 & 45.63 \\
& leetcode & 142 & 413,932 & 1,459 & 1,059 & 2 & 1 & 47.77 \\
& pony & 112 & 7,894 & 389 & 260 & 20 & 52 & 26.33 \\
& psychology & 101 & 52,835 & 693 & 384 & 7 & 1 & 45.95 \\
& robotics & 101 & 61,961 & 2,180 & 291 & 5 & 1 & 62.89 \\
& stackoverflow & 117 & 107,081 & 1,293 & 1,715 & 4 & 1 & 54.02 \\
& sustainable\_living & 108 & 60,792 & 683 & 344 & 5 & 1 & 47.94 \\
& theoremqa\_questions & 194 & 188,002 & 426 & 754 & 3 & 2 & 27.65 \\
& theoremqa\_theorems & 76 & 23,839 & 416 & 874 & 2 & 2 & 26.82 \\
\midrule
\multirow{4}{*}{\textbf{Multi-hop}} & 2wikimultihopqa & 12,576 & 125,237 & 68 & 377 & 2 & 1 & 6.34 \\
& musique & 2,417 & 48,315 & 102 & 524 & 3 & 1 & 8.64 \\
& novelhopqa & 4,345 & 4,345 & 138 & 2,336 & 1 & 1 & 11.64 \\
& hotpotqa & 7,405 & 5,233,329 & 92 & 289 & 2 & 1 & 8.60 \\
\bottomrule
\end{tabular}%
}
\caption{Statistics of evaluation benchmarks. Query Len and Doc Len are average character lengths. CW (Content Words) measures query complexity. HotpotQA appears in both BEIR and Multi-hop categories as they refer to the same dataset.}
\label{tab:dataset_stats}
\end{table*}

\section{Error Analysis on NovelHopQA}
\label{sec:appendix_error_analysis}

Table~\ref{tab:error_analysis} provides detailed error analysis comparing \textbf{Q/Doc=1} and \textbf{Q/Doc=3} predictions on NovelHopQA. Figure~\ref{fig:case_studies_appendix} shows additional case studies demonstrating the ``same-book-wrong-passage'' error pattern.

\begin{table}[htbp]
\centering
\small
\resizebox{\columnwidth}{!}{
\begin{tabular}{l|c}
\toprule
\textbf{Metric} & \textbf{Value} \\
\midrule
Total queries & 4,345 \\
\midrule
\textbf{Q/Doc=3} correct, \textbf{Q/Doc=1} wrong & 522 \\
\textbf{Q/Doc=1} correct, \textbf{Q/Doc=3} wrong & 187 \\
Net improvement (\textbf{Q/Doc=3} $-$ \textbf{Q/Doc=1}) & +335 \\
\midrule
\multicolumn{2}{l}{\textit{Among \textbf{Q/Doc=1} errors where \textbf{Q/Doc=3} succeeds:}} \\
\quad Same book, wrong passage & 455 (87.2\%) \\
\quad Different book & 67 (12.8\%) \\
\bottomrule
\end{tabular}
}
\caption{Error analysis comparing \textbf{Q/Doc=1} and \textbf{Q/Doc=3} on NovelHopQA (P@1) using Contriever. Statistics are computed on the \textbf{full test set} (4,345 queries across all 4 hop levels). The majority of \textbf{Q/Doc=1} errors retrieve the wrong passage from the correct book, indicating overfitting to surface features.}
\label{tab:error_analysis}
\end{table}

\begin{figure*}[!htbp]
\centering
\small

\begin{minipage}[t]{0.48\textwidth}
\begin{tcolorbox}[
  colback=white,
  colframe=gray!40,
  colbacktitle=gray!20,
  coltitle=black,
  title={\textbf{Query:} What is Leopold Bloom accused of being? \\ \textbf{Book:} \textit{Ulysses}},
  fonttitle=\small,
  boxrule=0.5pt,
  arc=4pt,
  left=4pt, right=4pt, top=4pt, bottom=4pt
]
\textbf{Q/Doc=1} \textcolor{red}{\ding{55}} \textit{``BLOOM: I have forgotten for the moment. Ah, yes! (He takes off his high grad...''} \\
\textcolor{gray}{$\hookrightarrow$ A dialogue mentioning Bloom} \\[0.5em]
\textbf{Q/Doc=3} \textcolor{green!50!black}{\ding{51}} \textit{``THE CRIER: (Loudly.) Whereas Leopold Bloom of no fixed abode is a wellknown dynamitard...''} \\
\textcolor{gray}{$\hookrightarrow$ The actual accusation against Bloom}
\end{tcolorbox}
\end{minipage}
\hfill
\begin{minipage}[t]{0.48\textwidth}
\begin{tcolorbox}[
  colback=white,
  colframe=gray!40,
  colbacktitle=gray!20,
  coltitle=black,
  title={\textbf{Query:} Why does Elinor suggest deferring the letter? \\ \textbf{Book:} \textit{Sense and Sensibility}},
  fonttitle=\small,
  boxrule=0.5pt,
  arc=4pt,
  left=4pt, right=4pt, top=4pt, bottom=4pt
]
\textbf{Q/Doc=1} \textcolor{red}{\ding{55}} \textit{``Elinor said no more. She was debating within herself on the eligibility of b...''} \\
\textcolor{gray}{$\hookrightarrow$ Elinor in a different context} \\[0.5em]
\textbf{Q/Doc=3} \textcolor{green!50!black}{\ding{51}} \textit{``As dinner was not to be ready in less than two hours from their arrival, Eli...''} \\
\textcolor{gray}{$\hookrightarrow$ Elinor's suggestion about the letter}
\end{tcolorbox}
\end{minipage}

\vspace{0.5em}

\begin{minipage}[t]{0.48\textwidth}
\begin{tcolorbox}[
  colback=white,
  colframe=gray!40,
  colbacktitle=gray!20,
  coltitle=black,
  title={\textbf{Query:} Who prevented Sancho from being robbed? \\ \textbf{Book:} \textit{Don Quixote}},
  fonttitle=\small,
  boxrule=0.5pt,
  arc=4pt,
  left=4pt, right=4pt, top=4pt, bottom=4pt
]
\textbf{Q/Doc=1} \textcolor{red}{\ding{55}} \textit{``Sancho came home in such glee and spirits that his wife noticed his happiness...''} \\
\textcolor{gray}{$\hookrightarrow$ Sancho in a different scene} \\[0.5em]
\textbf{Q/Doc=3} \textcolor{green!50!black}{\ding{51}} \textit{``And now day dawned; and if the dead freebooters had scared them, their hearts were no...''} \\
\textcolor{gray}{$\hookrightarrow$ The robbery prevention scene}
\end{tcolorbox}
\end{minipage}
\hfill
\begin{minipage}[t]{0.48\textwidth}
\begin{tcolorbox}[
  colback=white,
  colframe=gray!40,
  colbacktitle=gray!20,
  coltitle=black,
  title={\textbf{Query:} What is Emma questioning about Mr. Knightley? \\ \textbf{Book:} \textit{Emma}},
  fonttitle=\small,
  boxrule=0.5pt,
  arc=4pt,
  left=4pt, right=4pt, top=4pt, bottom=4pt
]
\textbf{Q/Doc=1} \textcolor{red}{\ding{55}} \textit{``Well, I believe, if you will excuse me, Mr. Knightley, if you will not consider...''} \\
\textcolor{gray}{$\hookrightarrow$ A dialogue with Mr. Knightley} \\[0.5em]
\textbf{Q/Doc=3} \textcolor{green!50!black}{\ding{51}} \textit{``Emma could not help laughing as she answered, `Upon my word, I believe you know her...''} \\
\textcolor{gray}{$\hookrightarrow$ Emma questioning Knightley's understanding}
\end{tcolorbox}
\end{minipage}

\caption{Additional Contriever case studies from NovelHopQA. In all cases, Q/Doc=1 retrieves passages that merely mention the character name, while Q/Doc=3 correctly identifies passages containing the specific semantic content requested by the query.}
\label{fig:case_studies_appendix}
\end{figure*}

\section{Optimal Q/Doc Distribution}
\label{sec:appendix_qdoc}

\begin{figure}[!htbp]
\centering
\begin{tikzpicture}
\begin{axis}[
    ybar,
    width=0.52\columnwidth,
    height=4cm,
    bar width=10pt,
    ylabel={\textbf{Datasets}},
    xlabel={Optimal Q/Doc},
    ymin=0,
    ymax=12,
    xtick=data,
    symbolic x coords={1,2,5,10,20},
    nodes near coords,
    nodes near coords align={vertical},
    title={\small\textbf{(a) Contriever}},
    title style={at={(0.5,1.02)}},
    enlarge x limits=0.12,
    every node near coord/.append style={font=\scriptsize\bfseries},
    label style={font=\small},
    tick label style={font=\footnotesize},
]
\addplot[fill=blue!60, draw=blue!80] coordinates {(1,10) (2,9) (5,4) (10,2) (20,6)};
\end{axis}
\end{tikzpicture}
\hspace{-0.5em}
\begin{tikzpicture}
\begin{axis}[
    ybar,
    width=0.52\columnwidth,
    height=4cm,
    bar width=10pt,
    ylabel={},
    yticklabels={},
    xlabel={Optimal Q/Doc},
    ymin=0,
    ymax=12,
    xtick=data,
    symbolic x coords={1,2,5,10,20},
    nodes near coords,
    nodes near coords align={vertical},
    title={\small\textbf{(b) RetroMAE}},
    title style={at={(0.5,1.02)}},
    enlarge x limits=0.12,
    every node near coord/.append style={font=\scriptsize\bfseries},
    label style={font=\small},
    tick label style={font=\footnotesize},
]
\addplot[fill=orange!60, draw=orange!80] coordinates {(1,7) (2,9) (5,3) (10,8) (20,4)};
\end{axis}
\end{tikzpicture}
\caption{Distribution of optimal Q/Doc ratios across 31 datasets. The optimal ratio varies from 1 to 20, with no universally optimal value.}
\label{fig:qdoc_distribution}
\end{figure}

\section{Per-Dataset Results for CW Weighting}
\label{sec:appendix_cw_full}

Table~\ref{tab:cw_weighting_full} provides the per-dataset NDCG@10 breakdown for the CW-weighted training experiment (Table~\ref{tab:cw_weighting} in the main text).

\begin{table*}[t]
\centering
\small
\begin{tabular}{l|cc|c}
\toprule
\textbf{Dataset} & \textbf{Standard} & \textbf{CW-weighted} & \textbf{$\Delta$} \\
\midrule
\multicolumn{4}{c}{\textit{TREC-DL}} \\
\midrule
TREC-DL 2019 & \textbf{56.99} & 55.15 & $-$1.84 \\
TREC-DL 2020 & \textbf{58.04} & 56.60 & $-$1.44 \\
\midrule
\textbf{Average} & \textbf{57.52} & 55.88 & $-$1.64 \\
\midrule
\multicolumn{4}{c}{\textit{BEIR}} \\
\midrule
TREC-COVID & 45.53 & \textbf{45.89} & +0.36 \\
NFCorpus & 31.38 & \textbf{31.66} & +0.28 \\
NQ & \textbf{34.50} & 34.53 & +0.03 \\
HotpotQA & 52.01 & \textbf{52.79} & +0.78 \\
FiQA & 26.34 & \textbf{27.57} & +1.23 \\
ArguAna & 49.04 & \textbf{51.17} & +2.13 \\
Touche-2020 & 18.01 & \textbf{18.23} & +0.22 \\
CQADupStack & 29.51 & \textbf{30.49} & +0.98 \\
Quora & \textbf{84.45} & 84.39 & $-$0.06 \\
DBPedia & 34.62 & \textbf{35.23} & +0.61 \\
SCIDOCS & 15.10 & \textbf{16.13} & +1.03 \\
FEVER & 66.01 & \textbf{67.62} & +1.61 \\
Climate-FEVER & 18.84 & \textbf{21.12} & +2.28 \\
SciFact & 60.06 & \textbf{62.42} & +2.36 \\
\midrule
\textbf{Average} & 40.39 & \textbf{41.38} & +0.99 \\
\midrule
\multicolumn{4}{c}{\textit{BRIGHT}} \\
\midrule
Biology & 5.92 & \textbf{7.05} & +1.13 \\
Earth Science & 11.11 & \textbf{12.48} & +1.37 \\
Economics & 9.55 & \textbf{10.29} & +0.74 \\
Psychology & 8.74 & \textbf{9.57} & +0.83 \\
Robotics & \textbf{6.59} & \textbf{6.59} & 0.00 \\
StackOverflow & 7.31 & \textbf{8.10} & +0.79 \\
Sustainable Living & \textbf{8.40} & 7.64 & $-$0.76 \\
LeetCode & \textbf{12.35} & \textbf{12.35} & 0.00 \\
Pony & \textbf{1.66} & 1.62 & $-$0.04 \\
AOPS & 4.22 & \textbf{4.79} & +0.57 \\
TheoremQA-Q & 6.92 & \textbf{7.85} & +0.93 \\
TheoremQA-T & 2.53 & \textbf{3.31} & +0.78 \\
\midrule
\textbf{Average} & 7.11 & \textbf{7.64} & +0.53 \\
\midrule
\multicolumn{4}{c}{\textit{Multi-hop}} \\
\midrule
HotpotQA & 52.01 & \textbf{52.86} & +0.85 \\
2WikiMultihopQA & 61.39 & \textbf{62.25} & +0.86 \\
MuSiQue & \textbf{33.72} & 33.21 & $-$0.51 \\
NovelHopQA & 54.53 & \textbf{58.68} & +4.15 \\
\midrule
\textbf{Average} & 50.41 & \textbf{51.75} & +1.34 \\
\bottomrule
\end{tabular}
\caption{Per-dataset NDCG@10 breakdown for CW-weighted training on Contriever with supervised MS MARCO data (80k single-query pairs). Standard: uniform sample weights. CW-weighted: content-word-based sample weights. CW weighting consistently improves BEIR (+0.99) and Multi-hop (+1.34) performance with slight TREC-DL trade-off ($-$1.64). Best scores in \textbf{bold}.}
\label{tab:cw_weighting_full}
\end{table*}

\section{Per-Dataset Results for ReasonEmbed Validation}
\label{sec:appendix_reasonembed_full}

Table~\ref{tab:reasonembed_full} provides the per-dataset NDCG@10 breakdown for all ReasonEmbed validation experiments (Table~\ref{tab:reasonembed} in the main text).

\begin{table*}[t]
\centering
\small
\resizebox{\textwidth}{!}{
\begin{tabular}{l|cc|cccc|cccc|cc}
\toprule
& \multicolumn{10}{c|}{\textit{w/ reasoning query}} & \multicolumn{2}{c}{\textit{w/o reasoning}} \\
\cmidrule(lr){2-11} \cmidrule(lr){12-13}
& \multicolumn{2}{c|}{query-only} & \multicolumn{4}{c|}{reasoning-only} & \multicolumn{4}{c|}{multi-query} & \multicolumn{2}{c}{query-only} \\
\cmidrule(lr){2-3} \cmidrule(lr){4-7} \cmidrule(lr){8-11} \cmidrule(lr){12-13}
\textbf{Dataset} & \textbf{RI} & \textbf{RI$\times$CW} & \textbf{--} & \textbf{RI} & \textbf{CW} & \textbf{RI$\times$CW} & \textbf{--} & \textbf{RI} & \textbf{CW} & \textbf{RI$\times$CW} & \textbf{--} & \textbf{CW} \\
& (ReasonEmbed) & & & & & & & & & & & \\
\midrule
\multicolumn{13}{c}{\textit{BRIGHT (in-domain)}} \\
\midrule
Biology & 13.49 & 13.80 & 8.28 & 13.49 & 8.47 & 13.23 & \textbf{16.16} & \underline{15.47} & 13.15 & 14.64 & \textbf{13.81} & 12.98 \\
Earth Science & \textbf{30.36} & \underline{29.31} & 16.92 & 23.11 & 16.23 & 22.85 & 27.64 & 28.78 & 25.05 & 27.60 & 26.49 & \textbf{26.79} \\
Economics & \textbf{22.39} & 20.76 & 18.13 & 20.53 & 17.21 & 19.31 & \underline{22.25} & 21.43 & 19.97 & 20.64 & 20.60 & \textbf{20.88} \\
Psychology & \textbf{24.29} & \underline{23.48} & 15.98 & 19.34 & 15.34 & 18.23 & 23.05 & 23.19 & 19.85 & 22.20 & 23.27 & \textbf{23.94} \\
Robotics & 13.03 & \textbf{13.44} & 10.36 & 10.17 & 11.15 & 9.67 & 12.66 & \underline{13.21} & 12.19 & 12.29 & \textbf{12.42} & 11.89 \\
StackOverflow & \underline{19.66} & 18.12 & 14.03 & 16.54 & 13.90 & 16.34 & 19.39 & \textbf{20.23} & 17.83 & 18.81 & 15.85 & \textbf{17.05} \\
Sustainable Living & 16.76 & 16.50 & 11.06 & 13.88 & 9.89 & 12.66 & \textbf{17.68} & \underline{16.97} & 16.63 & 16.69 & 15.85 & \textbf{17.31} \\
LeetCode & 9.12 & 9.04 & 8.90 & 7.64 & 8.23 & 6.62 & 8.19 & 9.07 & \textbf{10.13} & \underline{9.35} & 8.19 & \textbf{9.75} \\
Pony & \underline{1.38} & \textbf{1.42} & 0.90 & 0.48 & 0.75 & 0.55 & 0.55 & 0.95 & 0.99 & 0.89 & 1.01 & \textbf{1.32} \\
AOPS & 4.06 & 4.41 & 3.35 & 3.74 & 3.11 & 3.80 & 4.11 & \textbf{4.70} & \underline{4.46} & 4.43 & \textbf{3.11} & 2.71 \\
TheoremQA-Q & \textbf{26.97} & \underline{25.58} & 18.18 & 22.15 & 18.73 & 20.27 & 19.84 & 22.48 & 20.83 & 20.18 & \textbf{23.15} & 21.34 \\
TheoremQA-T & \textbf{30.88} & \underline{28.83} & 21.44 & 23.63 & 21.54 & 23.53 & 25.21 & 26.12 & 25.56 & 26.27 & 27.68 & \textbf{29.71} \\
\midrule
\textbf{BRIGHT Avg} & \textbf{17.70} & \underline{17.06} & 12.29 & 14.56 & 12.05 & 13.92 & 16.39 & 16.88 & 15.55 & 16.17 & 15.95 & \textbf{16.31} \\
\midrule
\multicolumn{13}{c}{\textit{Multi-hop (OOD)}} \\
\midrule
HotpotQA & 14.65 & \underline{15.75} & 5.95 & 15.59 & 5.34 & 11.71 & 12.47 & \textbf{16.28} & 15.04 & 14.93 & 12.35 & \textbf{12.38} \\
2WikiMultihopQA & 13.03 & 17.31 & 5.30 & 14.04 & 4.64 & 9.85 & 12.47 & \underline{18.79} & 18.77 & \textbf{19.22} & 10.60 & \textbf{13.03} \\
MuSiQue & 16.60 & 18.50 & 10.30 & 18.52 & 9.55 & 15.89 & 17.28 & \underline{19.37} & \textbf{19.51} & 19.27 & 15.54 & \textbf{17.32} \\
NovelHopQA & 75.09 & 77.05 & 63.44 & \underline{78.28} & 62.40 & 77.49 & 76.03 & \textbf{78.38} & 75.55 & 77.16 & \textbf{72.54} & 72.33 \\
\midrule
\textbf{Multi-hop Avg} & 29.84 & 32.15 & 21.25 & 31.61 & 20.48 & 28.74 & 29.56 & \textbf{33.20} & 32.22 & \underline{32.64} & 27.76 & \textbf{28.76} \\
\bottomrule
\end{tabular}}
\caption{Per-dataset NDCG@10 breakdown for ReasonEmbed validation (8k samples) with Qwen3-Embedding-0.6B. Methods organized by: (1) reasoning query usage, (2) training data type, (3) weighting scheme (--, RI, CW, RI$\times$CW). Best in \textbf{bold}, second-best \underline{underlined} within each group.}
\label{tab:reasonembed_full}
\end{table*}

\section{Full Diversity Ablation Results}
\label{sec:appendix_diversity_ablation}

Table~\ref{tab:diversity_ablation} provides the complete results comparing Paraphrase and Diverse training across all benchmark types, complementing Table~\ref{tab:multihop_diversity} in the main text. Figure~\ref{fig:diversity_ablation_qd} visualizes the Q-D metrics for these configurations.

\begin{figure}[!htbp]
\centering
\begin{tikzpicture}
\node[anchor=south] at (1.0, 2.2) {\textbf{(a) Quality $\uparrow$}};

\begin{axis}[
    name=plot1,
    width=0.50\columnwidth,
    height=0.47\columnwidth,
    ylabel={\textbf{Dist-Sim}},
    xmin=0.5, xmax=3.5,
    ymin=0.55, ymax=0.80,
    xtick={1, 2, 3},
    xticklabels={},
    ytick={0.6, 0.7, 0.8},
    tick label style={font=\scriptsize},
    label style={font=\small},
    grid=major,
]
\addplot[color=green!70!black, mark=triangle*, thick] coordinates {(1, 0.73) (2, 0.72) (3, 0.72)};
\addplot[color=blue, mark=*, thick, line width=1.2pt] coordinates {(1, 0.68) (2, 0.65) (3, 0.63)};
\end{axis}

\begin{axis}[
    at={(plot1.south)},
    anchor=north,
    yshift=-0.5cm,
    name=plot2,
    width=0.50\columnwidth,
    height=0.47\columnwidth,
    xlabel={Queries/Doc},
    ylabel={\textbf{Len-Sim}},
    xmin=0.5, xmax=3.5,
    ymin=0.50, ymax=0.70,
    xtick={1, 2, 3},
    xticklabels={5, 10, 20},
    ytick={0.55, 0.60, 0.65},
    tick label style={font=\scriptsize},
    label style={font=\small},
    grid=major,
]
\addplot[color=green!70!black, mark=triangle*, thick] coordinates {(1, 0.59) (2, 0.58) (3, 0.57)};
\addplot[color=blue, mark=*, thick, line width=1.2pt] coordinates {(1, 0.62) (2, 0.59) (3, 0.60)};
\end{axis}

\node[anchor=south] at (5.0, 2.2) {\textbf{(b) Diversity $\downarrow$}};

\begin{axis}[
    at={(plot1.east)},
    anchor=west,
    xshift=1.8cm,
    name=plot3,
    width=0.50\columnwidth,
    height=0.47\columnwidth,
    ylabel={\textbf{CE}},
    xmin=0.5, xmax=3.5,
    ymin=0, ymax=0.65,
    xtick={1, 2, 3},
    xticklabels={},
    ytick={0, 0.2, 0.4, 0.6},
    tick label style={font=\scriptsize},
    label style={font=\small},
    grid=major,
]
\addplot[color=green!70!black, mark=triangle*, thick] coordinates {(1, 0.55) (2, 0.48) (3, 0.45)};
\addplot[color=blue, mark=*, thick, line width=1.2pt] coordinates {(1, 0.04) (2, 0.03) (3, 0.04)};
\end{axis}

\begin{axis}[
    at={(plot3.south)},
    anchor=north,
    yshift=-0.5cm,
    name=plot4,
    width=0.50\columnwidth,
    height=0.47\columnwidth,
    xlabel={Queries/Doc},
    ylabel={\textbf{Self-BLEU}},
    xmin=0.5, xmax=3.5,
    ymin=0, ymax=0.45,
    xtick={1, 2, 3},
    xticklabels={5, 10, 20},
    ytick={0, 0.1, 0.2, 0.3, 0.4},
    tick label style={font=\scriptsize},
    label style={font=\small},
    grid=major,
]
\addplot[color=green!70!black, mark=triangle*, thick] coordinates {(1, 0.21) (2, 0.30) (3, 0.35)};
\addplot[color=blue, mark=*, thick, line width=1.2pt] coordinates {(1, 0.15) (2, 0.18) (3, 0.19)};
\end{axis}

\node[anchor=base] at (0.8, -4.0) {\footnotesize\raisebox{0pt}{\textcolor{green!70!black}{$\blacktriangle$}} Paraphrase};
\node[anchor=base] at (3.8, -4.0) {\footnotesize\raisebox{-1pt}{\textcolor{blue}{$\bullet$}} Diverse};
\end{tikzpicture}
\caption{Quality ($\uparrow$) and Diversity ($\downarrow$) metrics for Paraphrase vs Diverse variants (trained on 8k documents). Higher quality values indicate more similarity to human-annotated queries; lower diversity values indicate higher diversity. Both maintain similar quality, but Diverse achieves dramatically lower CE and Self-BLEU.}
\label{fig:diversity_ablation_qd}
\end{figure}

\begin{table*}[!htbp]
\centering
\small
\resizebox{\textwidth}{!}{
\begin{tabular}{l|c|cccc|cccc}
\toprule
& & \multicolumn{4}{c|}{\textbf{Contriever}} & \multicolumn{4}{c}{\textbf{RetroMAE}} \\
\textbf{Variant} & \textbf{Q/Doc} & \textbf{TREC-DL (2)} & \textbf{BEIR (14)} & \textbf{BRIGHT (12)} & \textbf{Multi-hop (4)} & \textbf{TREC-DL (2)} & \textbf{BEIR (14)} & \textbf{BRIGHT (12)} & \textbf{Multi-hop (4)} \\
\midrule
Paraphrase & 5 & \textbf{54.89} & \textbf{41.15} & 7.89 & 51.60 & \textbf{54.34} & \textbf{39.67} & 7.15 & 51.97 \\
Diverse & 5 & 48.86 & 40.70 & \textbf{9.14} & \textbf{54.40} & 49.37 & 38.85 & \textbf{8.02} & \textbf{56.35} \\
\midrule
Paraphrase & 10 & \textbf{52.99} & \textbf{40.36} & 8.22 & 52.53 & \textbf{51.98} & \textbf{38.81} & 8.01 & 52.02 \\
Diverse & 10 & 47.22 & 39.56 & \textbf{9.26} & \textbf{53.88} & 46.61 & 37.91 & \textbf{8.25} & \textbf{55.91} \\
\midrule
Paraphrase & 20 & \textbf{53.72} & \textbf{40.08} & 8.23 & 51.06 & \textbf{52.30} & \textbf{38.59} & \textbf{8.00} & 52.06 \\
Diverse & 20 & 44.37 & 38.35 & \textbf{9.30} & \textbf{52.52} & 43.27 & 36.46 & 7.86 & \textbf{52.49} \\
\bottomrule
\end{tabular}}
\caption{Effect of diversity level at different query counts, trained on 8k documents. Paraphrase generates semantically similar queries (CE$\approx$0.50), while Diverse generates varied queries (CE$\approx$0.04). Both backbones show consistent patterns: Paraphrase performs better on in-domain (TREC-DL) and standard OOD (BEIR), while Diverse excels on reasoning-intensive (BRIGHT) and multi-hop tasks.}
\label{tab:diversity_ablation}
\end{table*}

\section{Full Query Scaling Results}
\label{sec:appendix_query_scaling}

Table~\ref{tab:scaling_full} provides detailed query scaling results for TREC-DL, BEIR, and BRIGHT benchmarks, complementing Table~\ref{tab:multihop_scaling} in the main text. Table~\ref{tab:query_scaling} provides a summary across all benchmark types.

\begin{table*}[!htbp]
\centering
\small
\begin{tabular}{l|ccccc|ccccc}
\toprule
& \multicolumn{5}{c|}{\textbf{Contriever}} & \multicolumn{5}{c}{\textbf{RetroMAE}} \\
\textbf{Q/Doc} & \textbf{1} & \textbf{2} & \textbf{5} & \textbf{10} & \textbf{20} & \textbf{1} & \textbf{2} & \textbf{5} & \textbf{10} & \textbf{20} \\
\midrule
\multicolumn{11}{c}{\textit{TREC-DL}} \\
\midrule
DL-19 & \textbf{52.46} & 50.66 & 48.57 & 46.80 & 44.93 & 54.04 & \textbf{55.93} & 52.76 & 49.50 & 45.86 \\
DL-20 & \textbf{50.55} & 50.01 & 49.16 & 47.64 & 43.81 & 46.99 & \textbf{46.60} & 45.98 & 43.73 & 40.68 \\
\midrule
\multicolumn{11}{c}{\textit{BEIR}} \\
\midrule
ArguAna & 44.04 & 50.46 & 55.18 & \textbf{56.46} & 56.12 & 40.59 & 46.85 & 52.08 & 53.01 & \textbf{53.22} \\
C-FEVER & 21.25 & 26.39 & \textbf{27.21} & 25.31 & 23.97 & 23.28 & \textbf{25.15} & 25.14 & 23.86 & 22.79 \\
CQADup & \textbf{29.75} & 29.92 & 29.16 & 28.19 & 27.77 & 25.37 & \textbf{26.69} & 26.42 & 26.23 & 25.42 \\
DBPedia & 33.39 & 33.45 & \textbf{33.91} & 33.61 & 33.29 & 31.89 & \textbf{32.18} & 31.59 & 30.46 & 29.00 \\
FEVER & \textbf{67.17} & 66.71 & 65.39 & 60.93 & 55.03 & \textbf{71.28} & 68.85 & 64.84 & 58.06 & 52.38 \\
FiQA & 24.59 & \textbf{26.03} & 25.92 & 25.34 & 24.37 & 19.48 & 21.44 & 22.03 & \textbf{22.77} & 21.54 \\
NFCorpus & 31.59 & \textbf{32.86} & 32.55 & 32.10 & 32.04 & 27.25 & 28.74 & \textbf{28.97} & 28.93 & 28.78 \\
NQ & 29.24 & \textbf{30.30} & 29.28 & 28.22 & 24.25 & 28.32 & \textbf{29.57} & 28.96 & 27.88 & 24.02 \\
Quora & \textbf{83.68} & 83.19 & 81.95 & 80.90 & 80.39 & \textbf{81.82} & 81.40 & 80.16 & 79.40 & 77.95 \\
SCIDOCS & 17.29 & \textbf{17.61} & 17.23 & 16.88 & 16.27 & 14.22 & 14.64 & \textbf{14.76} & 14.16 & 13.98 \\
SciFact & 59.48 & 64.39 & 66.02 & 66.06 & \textbf{67.56} & 50.34 & 55.66 & 58.35 & 58.69 & \textbf{60.32} \\
T-COVID & \textbf{38.01} & 36.14 & 35.12 & 31.66 & 30.88 & \textbf{44.40} & 41.61 & 40.20 & 39.56 & 39.06 \\
Touche & \textbf{18.91} & 17.25 & 14.35 & 12.57 & 10.09 & \textbf{19.94} & 18.90 & 17.47 & 16.29 & 13.39 \\
\midrule
\multicolumn{11}{c}{\textit{BRIGHT}} \\
\midrule
AOPS & 2.86 & 3.29 & 3.80 & \textbf{4.68} & 4.51 & 0.93 & 2.78 & 3.43 & \textbf{3.73} & 3.26 \\
Biology & 6.45 & \textbf{9.38} & 8.32 & 8.73 & 7.93 & 4.81 & 6.52 & 7.89 & \textbf{8.11} & 6.67 \\
Earth Sci. & 11.90 & 17.13 & \textbf{19.13} & 18.60 & 18.76 & 13.52 & 13.92 & \textbf{17.14} & 16.71 & 16.32 \\
Econ. & 9.12 & \textbf{10.70} & 9.47 & 9.67 & 8.33 & \textbf{10.61} & 8.00 & 8.86 & 8.76 & 7.54 \\
LeetCode & 12.19 & 13.33 & 14.02 & 14.64 & \textbf{14.49} & 12.39 & 13.65 & 14.80 & \textbf{14.79} & 14.64 \\
Pony & 6.78 & \textbf{10.28} & 6.48 & 6.76 & 7.78 & \textbf{8.66} & 7.12 & 4.86 & 5.05 & 5.00 \\
Psych. & \textbf{13.43} & 12.99 & 11.28 & 11.10 & 11.51 & 10.75 & 11.03 & 10.61 & \textbf{11.47} & 10.93 \\
Robotics & 5.34 & 7.58 & 7.68 & 7.92 & \textbf{8.97} & 7.11 & 8.06 & 9.29 & \textbf{9.84} & 8.48 \\
SO & 7.11 & \textbf{9.63} & 8.81 & 8.23 & 8.54 & 5.81 & 5.65 & 5.04 & 5.00 & \textbf{6.80} \\
Sustain. & \textbf{7.89} & 8.00 & 6.80 & 6.31 & 4.72 & \textbf{8.95} & 6.70 & 6.51 & 6.18 & 5.39 \\
TQA-Q & 5.91 & 7.64 & 8.70 & 9.14 & \textbf{9.67} & 2.95 & 4.05 & 5.34 & 5.34 & \textbf{5.72} \\
TQA-T & 1.65 & 4.10 & 5.13 & 5.39 & \textbf{6.42} & 1.45 & 1.35 & 2.43 & \textbf{4.02} & 3.54 \\
\bottomrule
\end{tabular}
\caption{NDCG@10 (\%) when scaling diverse queries per document (8k documents). Column headers indicate Q/Doc values (1, 2, 5, 10, 20). DL-19/20=TREC-DL 2019/2020, C-FEVER=Climate-FEVER, CQADup=CQADupStack, T-COVID=TREC-COVID, Touche=Touche-2020, Earth Sci.=Earth Science, Econ.=Economics, Psych.=Psychology, SO=StackOverflow, Sustain.=Sustainable Living, TQA-Q/T=TheoremQA Questions/Theorems. In-domain tasks (TREC-DL) peak at Q/Doc=1-2, while reasoning-intensive tasks (BRIGHT) often benefit from more queries.}
\label{tab:scaling_full}
\end{table*}

\begin{table*}[!htbp]
\centering
\small
\resizebox{\textwidth}{!}{
\begin{tabular}{c|cccc|cccc}
\toprule
& \multicolumn{4}{c|}{\textbf{Contriever}} & \multicolumn{4}{c}{\textbf{RetroMAE}} \\
\textbf{Q/Doc} & \textbf{TREC-DL (2)} & \textbf{BEIR (14)} & \textbf{BRIGHT (12)} & \textbf{Multi-hop (4)} & \textbf{TREC-DL (2)} & \textbf{BEIR (14)} & \textbf{BRIGHT (12)} & \textbf{Multi-hop (4)} \\
\midrule
1 & \textbf{51.50} & 39.41 & 7.55 & 52.33 & 50.52 & 37.88 & 7.33 & 51.56 \\
2 & 50.33 & \textbf{40.72} & \textbf{9.50} & 54.34 & \textbf{51.26} & \textbf{38.97} & 7.40 & \textbf{56.91} \\
5 & 48.86 & 40.70 & 9.14 & \textbf{54.40} & 49.37 & 38.85 & 8.02 & 56.35 \\
10 & 47.22 & 39.56 & 9.26 & 53.88 & 46.61 & 37.91 & \textbf{8.25} & 55.91 \\
20 & 44.37 & 38.35 & 9.30 & 52.52 & 43.27 & 36.46 & 7.86 & 52.49 \\
\bottomrule
\end{tabular}}
\caption{Effect of scaling diverse queries per document (8k documents). Performance peaks around Q/Doc=2-5 for most OOD benchmarks, then declines with more queries. Note that this optimal range is relative to the 8k document subset; larger corpora may require more queries per document to achieve optimal coverage. Best results per column are in \textbf{bold}.}
\label{tab:query_scaling}
\end{table*}

\section{Cost Efficiency Analysis}
\label{sec:appendix_cost}

Table~\ref{tab:cost_efficiency_full} provides full results for the cost efficiency experiment shown in Figure~\ref{fig:cost_efficiency_multihop}. All configurations use 80k total training pairs but vary the document-query ratio.

Figures~\ref{fig:cost_efficiency_full} and \ref{fig:cost_efficiency_full_retromae} show results across all four benchmark types (TREC-DL, BEIR, BRIGHT, Multi-hop), comparing our method with baselines that use 80k documents. The pattern differs by task type: TREC-DL and BEIR prefer document coverage, while BRIGHT and Multi-hop tolerate or benefit from reduced documents with increased query diversity.

\begin{table*}[htbp]
\centering
\small
\resizebox{\textwidth}{!}{
\begin{tabular}{l|r|r|cccc|cccc}
\toprule
& & & \multicolumn{4}{c|}{\textbf{Contriever}} & \multicolumn{4}{c}{\textbf{RetroMAE}} \\
\textbf{Docs} & \textbf{Q/Doc} & \textbf{LLM Cost} & \textbf{TREC-DL (2)} & \textbf{BEIR (14)} & \textbf{BRIGHT (12)} & \textbf{Multi-hop (4)} & \textbf{TREC-DL (2)} & \textbf{BEIR (14)} & \textbf{BRIGHT (12)} & \textbf{Multi-hop (4)} \\
\midrule
80k & 1 & 100\% & \textbf{52.94} & \textbf{41.43} & 8.57 & 52.77 & \textbf{54.00} & 39.31 & 7.72 & 54.56 \\
40k & 2 & 50\% & 52.04 & 40.89 & 8.89 & \textbf{53.97} & 53.88 & \textbf{39.55} & 7.73 & \textbf{55.79} \\
16k & 5 & 20\% & 48.67 & 41.14 & \textbf{9.32} & 52.30 & 50.03 & 39.45 & 7.89 & 54.23 \\
8k & 10 & 10\% & 47.22 & 39.56 & 9.26 & 53.88 & 46.61 & 37.91 & \textbf{8.25} & 55.91 \\
4k & 20 & 5\% & 45.66 & 38.18 & 9.14 & 51.49 & 44.44 & 36.86 & 8.05 & 51.88 \\
\bottomrule
\end{tabular}}
\caption{Efficiency analysis with fixed 80k training pairs. LLM cost is proportional to the number of documents (not queries per document), since multiple queries can be generated in a single API call. Reducing documents from 80k to 4k (5\% LLM cost) maintains competitive OOD performance, especially on BRIGHT.}
\label{tab:cost_efficiency_full}
\end{table*}

\begin{figure}[!htbp]
\centering
\begin{tikzpicture}
\begin{axis}[
    name=plot1,
    width=0.56\columnwidth,
    height=0.58\columnwidth,
    title={\textbf{TREC-DL (2)}},
    title style={font=\small},
    ylabel={\textbf{NDCG@10}},
    xmode=log,
    log basis x=2,
    xmin=3, xmax=100,
    x dir=reverse,
    ymin=44, ymax=58,
    xtick={4, 8, 16, 40, 80},
    xticklabels={},
    ytick={46, 50, 54, 58},
    tick label style={font=\footnotesize},
    label style={font=\small},
    grid=major,
]
\addplot[color=blue, mark=*, thick, line width=1pt] coordinates {
    (80, 52.94) (40, 52.04) (16, 48.67) (8, 47.22) (4, 45.66)
};
\addplot[color=magenta!80!black, mark=pentagon*, thick, only marks, mark size=2.5pt] coordinates {(80, 57.52)};
\addplot[color=red, mark=square*, thick, only marks, mark size=2.5pt] coordinates {(80, 56.33)};
\addplot[color=orange, mark=triangle*, thick, only marks, mark size=2.5pt] coordinates {(80, 53.53)};
\addplot[color=purple, mark=diamond*, thick, only marks, mark size=2.5pt] coordinates {(80, 49.61)};
\end{axis}

\begin{axis}[
    at={(plot1.east)},
    anchor=west,
    xshift=0.5cm,
    name=plot2,
    width=0.56\columnwidth,
    height=0.58\columnwidth,
    title={\textbf{BRIGHT (12)}},
    title style={font=\small},
    xmode=log,
    log basis x=2,
    xmin=3, xmax=100,
    x dir=reverse,
    ymin=7, ymax=10,
    xtick={4, 8, 16, 40, 80},
    xticklabels={},
    ytick={7, 8, 9, 10},
    tick label style={font=\footnotesize},
    label style={font=\small},
    grid=major,
]
\addplot[color=blue, mark=*, thick, line width=1pt] coordinates {
    (80, 8.57) (40, 8.89) (16, 9.32) (8, 9.26) (4, 9.14)
};
\addplot[color=magenta!80!black, mark=pentagon*, thick, only marks, mark size=2.5pt] coordinates {(80, 7.11)};
\addplot[color=red, mark=square*, thick, only marks, mark size=2.5pt] coordinates {(80, 7.84)};
\addplot[color=orange, mark=triangle*, thick, only marks, mark size=2.5pt] coordinates {(80, 9.49)};
\addplot[color=purple, mark=diamond*, thick, only marks, mark size=2.5pt] coordinates {(80, 9.43)};
\end{axis}

\begin{axis}[
    at={(plot1.south)},
    anchor=north,
    yshift=-0.9cm,
    name=plot3,
    width=0.56\columnwidth,
    height=0.58\columnwidth,
    title={\textbf{BEIR (14)}},
    title style={font=\small},
    xlabel={\#Docs (k)},
    ylabel={\textbf{NDCG@10}},
    xmode=log,
    log basis x=2,
    xmin=3, xmax=100,
    x dir=reverse,
    ymin=36, ymax=43,
    xtick={4, 8, 16, 40, 80},
    xticklabels={4, 8, 16, 40, 80},
    ytick={36, 38, 40, 42},
    tick label style={font=\footnotesize},
    label style={font=\small},
    grid=major,
]
\addplot[color=blue, mark=*, thick, line width=1pt] coordinates {
    (80, 41.43) (40, 40.89) (16, 41.14) (8, 39.56) (4, 38.18)
};
\addplot[color=magenta!80!black, mark=pentagon*, thick, only marks, mark size=2.5pt] coordinates {(80, 40.39)};
\addplot[color=red, mark=square*, thick, only marks, mark size=2.5pt] coordinates {(80, 41.31)};
\addplot[color=orange, mark=triangle*, thick, only marks, mark size=2.5pt] coordinates {(80, 41.37)};
\addplot[color=purple, mark=diamond*, thick, only marks, mark size=2.5pt] coordinates {(80, 37.36)};
\end{axis}

\begin{axis}[
    at={(plot2.south)},
    anchor=north,
    yshift=-0.9cm,
    name=plot4,
    width=0.56\columnwidth,
    height=0.58\columnwidth,
    title={\textbf{Multi-hop (4)}},
    title style={font=\small},
    xlabel={\#Docs (k)},
    xmode=log,
    log basis x=2,
    xmin=3, xmax=100,
    x dir=reverse,
    ymin=49, ymax=55,
    xtick={4, 8, 16, 40, 80},
    xticklabels={4, 8, 16, 40, 80},
    ytick={50, 52, 54},
    tick label style={font=\footnotesize},
    label style={font=\small},
    grid=major,
]
\addplot[color=blue, mark=*, thick, line width=1pt] coordinates {
    (80, 52.77) (40, 53.97) (16, 52.30) (8, 53.88) (4, 51.49)
};
\addplot[color=magenta!80!black, mark=pentagon*, thick, only marks, mark size=2.5pt] coordinates {(80, 50.41)};
\addplot[color=red, mark=square*, thick, only marks, mark size=2.5pt] coordinates {(80, 52.72)};
\addplot[color=orange, mark=triangle*, thick, only marks, mark size=2.5pt] coordinates {(80, 52.58)};
\addplot[color=purple, mark=diamond*, thick, only marks, mark size=2.5pt] coordinates {(80, 50.72)};
\end{axis}

\node[anchor=base] at (0.0, -5.3) {\footnotesize\raisebox{1.5pt}{\textcolor{magenta!80!black}{\pgfuseplotmark{pentagon*}}} Sup.};
\node[anchor=base] at (1.1, -5.3) {\footnotesize\raisebox{0.5pt}{\textcolor{red}{$\blacksquare$}} InPars};
\node[anchor=base] at (2.3, -5.3) {\footnotesize\raisebox{0pt}{\textcolor{orange}{$\blacktriangle$}} SAP};
\node[anchor=base] at (3.6, -5.3) {\footnotesize\raisebox{0.5pt}{\textcolor{purple}{$\blacklozenge$}} DRAMA};
\node[anchor=base] at (4.8, -5.3) {\footnotesize\raisebox{-1pt}{\textcolor{blue}{$\bullet$}} Ours};
\end{tikzpicture}
\caption{Cost efficiency across all benchmark types with Contriever. We vary document-query configurations from 80k$\times$1 to 4k$\times$20 (x-axis shows document count). Baseline methods use 80k documents. TREC-DL and BEIR prefer higher document coverage, while BRIGHT and Multi-hop tolerate or benefit from reduced documents with increased query diversity.}
\label{fig:cost_efficiency_full}
\end{figure}

\begin{figure}[!htbp]
\centering
\begin{tikzpicture}
\begin{axis}[
    name=plot1,
    width=0.56\columnwidth,
    height=0.58\columnwidth,
    title={\textbf{TREC-DL (2)}},
    title style={font=\small},
    ylabel={\textbf{NDCG@10}},
    xmode=log,
    log basis x=2,
    xmin=3, xmax=100,
    x dir=reverse,
    ymin=43, ymax=58,
    xtick={4, 8, 16, 40, 80},
    xticklabels={},
    ytick={44, 48, 52, 56},
    tick label style={font=\footnotesize},
    label style={font=\small},
    grid=major,
]
\addplot[color=blue, mark=*, thick, line width=1pt] coordinates {
    (80, 54.00) (40, 53.88) (16, 50.03) (8, 46.61) (4, 44.44)
};
\addplot[color=magenta!80!black, mark=pentagon*, thick, only marks, mark size=2.5pt] coordinates {(80, 56.26)};
\addplot[color=red, mark=square*, thick, only marks, mark size=2.5pt] coordinates {(80, 56.07)};
\addplot[color=orange, mark=triangle*, thick, only marks, mark size=2.5pt] coordinates {(80, 54.53)};
\addplot[color=purple, mark=diamond*, thick, only marks, mark size=2.5pt] coordinates {(80, 52.12)};
\end{axis}

\begin{axis}[
    at={(plot1.east)},
    anchor=west,
    xshift=0.5cm,
    name=plot2,
    width=0.56\columnwidth,
    height=0.58\columnwidth,
    title={\textbf{BRIGHT (12)}},
    title style={font=\small},
    xmode=log,
    log basis x=2,
    xmin=3, xmax=100,
    x dir=reverse,
    ymin=6, ymax=9,
    xtick={4, 8, 16, 40, 80},
    xticklabels={},
    ytick={6, 7, 8, 9},
    tick label style={font=\footnotesize},
    label style={font=\small},
    grid=major,
]
\addplot[color=blue, mark=*, thick, line width=1pt] coordinates {
    (80, 7.72) (40, 7.73) (16, 7.89) (8, 8.25) (4, 8.05)
};
\addplot[color=magenta!80!black, mark=pentagon*, thick, only marks, mark size=2.5pt] coordinates {(80, 6.33)};
\addplot[color=red, mark=square*, thick, only marks, mark size=2.5pt] coordinates {(80, 6.98)};
\addplot[color=orange, mark=triangle*, thick, only marks, mark size=2.5pt] coordinates {(80, 7.43)};
\addplot[color=purple, mark=diamond*, thick, only marks, mark size=2.5pt] coordinates {(80, 6.42)};
\end{axis}

\begin{axis}[
    at={(plot1.south)},
    anchor=north,
    yshift=-0.9cm,
    name=plot3,
    width=0.56\columnwidth,
    height=0.58\columnwidth,
    title={\textbf{BEIR (14)}},
    title style={font=\small},
    xlabel={\#Docs (k)},
    ylabel={\textbf{NDCG@10}},
    xmode=log,
    log basis x=2,
    xmin=3, xmax=100,
    x dir=reverse,
    ymin=34, ymax=41,
    xtick={4, 8, 16, 40, 80},
    xticklabels={4, 8, 16, 40, 80},
    ytick={34, 36, 38, 40},
    tick label style={font=\footnotesize},
    label style={font=\small},
    grid=major,
]
\addplot[color=blue, mark=*, thick, line width=1pt] coordinates {
    (80, 39.31) (40, 39.55) (16, 39.45) (8, 37.91) (4, 36.86)
};
\addplot[color=magenta!80!black, mark=pentagon*, thick, only marks, mark size=2.5pt] coordinates {(80, 38.47)};
\addplot[color=red, mark=square*, thick, only marks, mark size=2.5pt] coordinates {(80, 39.02)};
\addplot[color=orange, mark=triangle*, thick, only marks, mark size=2.5pt] coordinates {(80, 39.69)};
\addplot[color=purple, mark=diamond*, thick, only marks, mark size=2.5pt] coordinates {(80, 35.38)};
\end{axis}

\begin{axis}[
    at={(plot2.south)},
    anchor=north,
    yshift=-0.9cm,
    name=plot4,
    width=0.56\columnwidth,
    height=0.58\columnwidth,
    title={\textbf{Multi-hop (4)}},
    title style={font=\small},
    xlabel={\#Docs (k)},
    xmode=log,
    log basis x=2,
    xmin=3, xmax=100,
    x dir=reverse,
    ymin=48, ymax=57,
    xtick={4, 8, 16, 40, 80},
    xticklabels={4, 8, 16, 40, 80},
    ytick={49, 52, 55},
    tick label style={font=\footnotesize},
    label style={font=\small},
    grid=major,
]
\addplot[color=blue, mark=*, thick, line width=1pt] coordinates {
    (80, 54.56) (40, 55.79) (16, 54.23) (8, 55.91) (4, 51.88)
};
\addplot[color=magenta!80!black, mark=pentagon*, thick, only marks, mark size=2.5pt] coordinates {(80, 51.15)};
\addplot[color=red, mark=square*, thick, only marks, mark size=2.5pt] coordinates {(80, 52.39)};
\addplot[color=orange, mark=triangle*, thick, only marks, mark size=2.5pt] coordinates {(80, 54.85)};
\addplot[color=purple, mark=diamond*, thick, only marks, mark size=2.5pt] coordinates {(80, 49.42)};
\end{axis}

\node[anchor=base] at (0.0, -5.3) {\footnotesize\raisebox{1.5pt}{\textcolor{magenta!80!black}{\pgfuseplotmark{pentagon*}}} Sup.};
\node[anchor=base] at (1.1, -5.3) {\footnotesize\raisebox{0.5pt}{\textcolor{red}{$\blacksquare$}} InPars};
\node[anchor=base] at (2.3, -5.3) {\footnotesize\raisebox{0pt}{\textcolor{orange}{$\blacktriangle$}} SAP};
\node[anchor=base] at (3.6, -5.3) {\footnotesize\raisebox{0.5pt}{\textcolor{purple}{$\blacklozenge$}} DRAMA};
\node[anchor=base] at (4.8, -5.3) {\footnotesize\raisebox{-1pt}{\textcolor{blue}{$\bullet$}} Ours};
\end{tikzpicture}
\caption{Cost efficiency with fixed 80k training pairs (RetroMAE backbone). Configuration and legend same as Figure~\ref{fig:cost_efficiency_full}. RetroMAE shows similar patterns: our method maintains competitive performance with fewer documents, particularly excelling on Multi-hop where 8k documents (10\% cost) achieves 55.91 NDCG@10, outperforming all 80k-document baselines.}
\label{fig:cost_efficiency_full_retromae}
\end{figure}

\section{Limitations of Few-shot Methods}
\label{sec:appendix_fewshot_limitations}

Existing methods like InPars \citep{bonifacio2022inpars} and SAP \citep{thakur2024swimir} employ few-shot prompting with carefully selected examples. While few-shot prompting can ensure high-quality queries that closely match target distributions, it has two critical limitations:

\paragraph{High Cost.} Few-shot methods require $M$ separate LLM calls to generate $M$ queries per document, and depend on powerful LLMs to follow complex few-shot patterns.

\paragraph{Limited Diversity.} LLMs tend to mimic the patterns of few-shot examples (syntax, length, question type), causing the output distribution to be anchored around these examples. Even with high temperature sampling ($>$0.7), the generated queries remain clustered around the exemplar patterns, producing only surface-level lexical variations rather than true semantic diversity. This results in high paraphrase ratios (CE$>$0.5), as shown in Table~\ref{tab:fewshot_vs_zeroshot}: few-shot queries for the same document are largely paraphrases of a single underlying pattern.

\section{Full Prompt Templates}
\label{sec:appendix_prompts}

Here we provide the complete prompt templates with all formatting details.

\begin{figure*}[!htbp]
\centering
\small
\begin{tcolorbox}[
  width=0.96\textwidth,
  colback=orange!8,
  colframe=orange!40,
  colbacktitle=orange!20,
  coltitle=black,
  title=\textbf{Paraphrase Prompt},
  fonttitle=\small,
  boxrule=0.5pt,
  arc=4pt,
  left=4pt, right=4pt, top=2pt, bottom=2pt
]
Your task is to generate \{M\} paraphrase queries based on the document(s). \\[0.3em]
\quad -- Identify \textbf{ONE main question} the document(s) answer \\
\quad -- Then rephrase it \{M\} different ways \\[0.3em]
All queries must ask the \underline{SAME question} with \underline{DIFFERENT wording}. \\[0.3em]
Document(s): \{document\} \\[0.3em]
Generate \{M\} queries: 1.
\end{tcolorbox}

\vspace{0.5em}

\begin{tcolorbox}[
  width=0.96\textwidth,
  colback=blue!5,
  colframe=blue!35,
  colbacktitle=blue!15,
  coltitle=black,
  title=\textbf{Diverse Prompt for Multi-hop Tasks},
  fonttitle=\small,
  boxrule=0.5pt,
  arc=4pt,
  left=4pt, right=4pt, top=2pt, bottom=2pt
]
Your task is to generate \{M\} independent queries based on the document(s). \\[0.3em]
You MUST generate queries in these specific formats: \\
\quad -- \textbf{What}... questions (factual) \\
\quad -- \textbf{How}... questions (procedural) \\
\quad -- \textbf{Why}... questions (causal) \\
\quad -- \textbf{When/If}... questions (conditional) \\
\quad -- \textbf{Keyword} queries (2-5 words, no question mark) \\
\quad -- \textbf{Statement/claim} format (e.g., ``X is used for Y'') \\
\quad -- \textbf{Which/Is it true}... questions \\
\quad -- \textbf{Comparison} or contrast questions \\[0.3em]
Each query must target \underline{different information} from the document. \\[0.3em]
Document(s): \{document\} \\[0.3em]
Generate \{M\} queries: 1.
\end{tcolorbox}
\caption{Full prompt templates. M denotes Q/Doc (the number of queries per document). To control experimental variables, when $M < 20$, we take the first M generated queries in order.}
\label{fig:full_prompts}
\end{figure*}

\section{Multilingual CW Configuration}
\label{sec:appendix_multilingual_cw}

This section provides implementation details for the multilingual experiments in Section~\ref{sec:multilingual}. Table~\ref{tab:multilingual_cw_config} summarizes the tokenization and stopword configuration.

\begin{table}[h]
\centering
\small
\begin{tabular}{lll}
\toprule
\textbf{Language} & \textbf{Tokenizer} & \textbf{Stopwords} \\
\midrule
Russian (ru) & Regex$^\dagger$ & stopwords-iso \\
Arabic (ar) & Regex$^\dagger$ & stopwords-iso \\
French (fr) & Regex$^\dagger$ & stopwords-iso \\
Chinese (zh) & jieba & stopwords-iso \\
Japanese (ja) & fugashi (MeCab) & stopwords-iso \\
Korean (ko) & kiwipiepy & stopwords-iso \\
\bottomrule
\end{tabular}
\caption{Multilingual CW computation configuration. $^\dagger$Regex tokenizer uses Unicode word boundary detection, sufficient for space-delimited languages. Stopwords are from the stopwords-iso collection (57 languages).}
\label{tab:multilingual_cw_config}
\end{table}

\paragraph{Design Rationale.} Space-delimited languages (Russian, Arabic, French) use regex tokenization. Chinese and Japanese require specialized segmenters (jieba and MeCab). Korean uses kiwipiepy for morphological analysis.

\paragraph{Stopword Observation.} We observe that standard French and Chinese stopword lists include common interrogative words, which may lead to CW underestimation. Table~\ref{tab:stopword_issue} shows examples. This could partially explain the degradation observed in Table~\ref{tab:miracl_results}, though other factors may also contribute.

\begin{table}[h]
\centering
\small
\begin{tabular}{lll}
\toprule
\textbf{Lang} & \textbf{Stopwords (question words)} & \textbf{Effect} \\
\midrule
fr & quel, comment, pourquoi, quand & CW$\leq$3 \\
zh & \begin{CJK}{UTF8}{gbsn}什么, 为什么, 哪里, 怎么\end{CJK} & CW$\leq$3 \\
\bottomrule
\end{tabular}
\caption{Interrogative words included in standard stopword lists, which may cause CW underestimation.}
\label{tab:stopword_issue}
\end{table}

\end{document}